\documentclass[%
 reprint,
 superscriptaddress,
 amsmath,amssymb,
 aps,
]{revtex4-2}

\usepackage{amsfonts}
\usepackage{amsmath}
\usepackage{algorithm}
\usepackage{algpseudocode}

\usepackage{graphicx}% Include figure files

\usepackage{dcolumn}% Align table columns on decimal point
\usepackage{bm}% bold math
\usepackage{hyperref}% add hypertext capabilities
\usepackage[caption=false]{subfig}

\usepackage{tikz}
\usetikzlibrary{arrows.meta, positioning, fit, calc, backgrounds}

\usepackage{color, soul}
\usepackage[usenames, dvipsnames]{xcolor} % xcolor provides more color names

\begin{document}
\setcitestyle{super}

\title{Observation geometry for uncertainty-aware Hamiltonian inference and experimental design in quantum magnets}

\author{Roy Liu}
\affiliation{%
 Group for AI in Materials Modeling and Analysis, The University of Texas at Austin, Austin, TX, USA.
}%
\affiliation{%
 Department of Computer Science, The University of Texas at Austin, Austin, TX, USA.
}%

\author{Venugopal Ranganathan}%
\affiliation{%
 Group for AI in Materials Modeling and Analysis, The University of Texas at Austin, Austin, TX, USA.
}%
\affiliation{%
 Walker Department of Mechanical Engineering, The University of Texas at Austin, Austin, TX, USA.
}%

\author{David Dahlbom}
\affiliation{
 Oak Ridge National Laboratory, Oak Ridge, TN, USA
}

\author{Shizhou Xu}%
\affiliation{
 Department of Mathematics, University of California Davis, Davis, CA, USA.
}
\affiliation{
 Stanford Institute for Materials and Energy Sciences, Stanford University, Stanford, CA 94305, USA
}

\author{Tianyu Zhang}%
\affiliation{%
 Group for AI in Materials Modeling and Analysis, The University of Texas at Austin, Austin, TX, USA.
}%
\affiliation{%
 Texas Materials Institute, The University of Texas at Austin, Austin, TX, USA.
}%

\author{Yuan Ni}
\affiliation{
 Stanford Institute for Materials and Energy Sciences, Stanford University, Stanford, CA 94305, USA
}
\affiliation{
 Linac Coherent Light Source, SLAC National Accelerator Laboratory, Menlo Park, CA, USA
}

\author{Daniel M.~Pajerowski}
\affiliation{
 Oak Ridge National Laboratory, Oak Ridge, TN, USA
}

\author{Garrett Granroth}
\affiliation{
 Oak Ridge National Laboratory, Oak Ridge, TN, USA
}

\author{Thomas Strohmer}
\affiliation{
 Department of Mathematics, University of California Davis, Davis, CA, USA.
}

\author{Matthew B.~Stone}
\affiliation{
 Oak Ridge National Laboratory, Oak Ridge, TN, USA
}

\author{Andrew F.~May}
\affiliation{
 Oak Ridge National Laboratory, Oak Ridge, TN, USA
}
\affiliation{
 Materials Science and Technology Division, Oak Ridge National Laboratory, Oak Ridge, TN, USA
}

\author{Mark D.~Lumsden}
\affiliation{
 Oak Ridge National Laboratory, Oak Ridge, TN, USA
}

\author{Joshua J.~Turner}
\affiliation{
 Stanford Institute for Materials and Energy Sciences, Stanford University, Stanford, CA 94305, USA
}
\affiliation{
 Linac Coherent Light Source, SLAC National Accelerator Laboratory, Menlo Park, CA, USA
}

\author{Yongqiang Cheng}
\email{chengy@ornl.gov}
\affiliation{
 Oak Ridge National Laboratory, Oak Ridge, TN, USA
}

\author{Zhantao Chen}
\email{zhantao@austin.utexas.edu}
\affiliation{%
 Group for AI in Materials Modeling and Analysis, The University of Texas at Austin, Austin, TX, USA.
}%
\affiliation{%
 Walker Department of Mechanical Engineering, The University of Texas at Austin, Austin, TX, USA.
}%
\affiliation{%
 Texas Materials Institute, The University of Texas at Austin, Austin, TX, USA.
}%

% \date{\today}

\begin{abstract}
Determining microscopic interactions from spectroscopic and scattering measurements is central to understanding quantum materials, yet it often remains unclear which interactions can be reliably revealed by the available experimental data and how additional experimental modalities should be designed to resolve the remaining ambiguities. Here we present an artificial intelligence-enabled framework for uncertainty-aware Hamiltonian inference and adaptive experimental design. By combining Hamiltonian-conditioned neural surrogates with Bayesian inference and observation geometry, the framework characterizes how measurements constrain Hamiltonian parameter space, quantifies the identifiability of microscopic interactions, and propagates posterior uncertainty directly in the physical Hamiltonian parameter space rather than an abstract learned representation. Using multimodal powder and single-crystal inelastic neutron scattering measurements of the quantum magnet NiPS\textsubscript{3}, we demonstrate physically interpretable Hamiltonian inference, modality-aware uncertainty quantification, and adaptive experimental design. The framework provides a general strategy for uncertainty-aware microscopic characterization and multimodal experimental design across quantum materials.
\end{abstract}


\maketitle

%\tableofcontents

% \section*{Todo}

% \begin{itemize}
%     \item add citations
%     \item update fig 4 if new data provided
% \end{itemize}

\section*{Introduction}

% \todo{Points needing highlights: preserve physics integrity; clear capability envelop, i.e., parameter-wise resolution; ...}

Microscopic models of elementary excitations provide a fundamental link between experimental observations and the underlying physics of condensed matter and materials. For quantum magnets, this connection can be modeled through the spin Hamiltonian, whose parameters encode exchange couplings, anisotropies, and other microscopic interactions governing the excitation spectrum~\cite{ntallis2021connection,petsch2023high,scheie2023spin,hase2024inelastic}. By fitting model parameters to measured spectra, one can infer these parameters, thereby determining the associated collective excitations as well as the static and dynamical properties of materials. In practice, however, this inverse problem is often ill-posed: distinct Hamiltonians can produce nearly indistinguishable spectra within experimental resolution, making it difficult to determine whether subtle microscopic features are genuinely constrained by the available measurements~\cite{gutenkunst2007universally,transtrum2015perspective,amari2016information,samarakoon2020machine,chen2023panoramic,samarakoon2022machine,misawa2026revisiting}. Consequently, conventional refinement procedures that report only a single best-fit Hamiltonian may obscure substantial parameter uncertainty, correlations, and alternative microscopic descriptions that remain consistent with the data~\cite{tarantola2005inverse,stuart2010inverse}. 
The central challenge is therefore not only to identify a Hamiltonian that reproduces the observed spectra, but also to determine which interactions are uniquely identifiable, which remain ambiguous, and how much microscopic information a given measurement can truly reveal.

Inelastic neutron scattering (INS) is a particularly powerful probe of magnetic and lattice excitations because it directly and quantitatively measures the momentum- and energy-resolved dynamical structure factor, $S(\mathbf{Q},\omega)$, for magnons and phonons with high resolution~\cite{squires1978introduction,lovesey1984theory,copley1993neutron,ivanov2022neutron,petsch2023high,scheie2023spin,hase2024inelastic,chatterji2005neutron}. Interpreting INS data typically relies on physically rigorous forward methods, including linear spin wave theory, exact diagonalization, density-matrix renormalization group calculations, and, more recently, neural quantum states, to generate theoretical spectra for comparison with experiment~\cite{jaklivc1994lanczos,schollwock2004density,toth2015linear,blosser2017finite,dahlbom2025sunny,carleo2017solving,mendes2023highly}. 
Although their computational scaling differs substantially, repeatedly evaluating these forward models across a high-dimensional Hamiltonian parameter space can remain prohibitively expensive.
Exhaustive grid-based searches scale exponentially with the dimensionality of the parameter space, whereas coarse parameter sweeps can overlook narrow regions containing plausible solutions.
Consequently, Hamiltonian refinement often relies on selected spectral cuts, expert-guided parameter tuning, or simplified Hamiltonian parameterizations, making it difficult to systematically quantify parameter uncertainty and determine which microscopic interactions are uniquely constrained by the experimental data.

Additional challenges arise from the increasingly multimodal nature of experimental materials characterization~\cite{chillal2020evidence,huang2024bimodal,vestin20264d,park20262d,moses2026cross}. In INS, both powder and single-crystal measurements are widely employed. Single-crystal INS preserves directional momentum information and can resolve fine features of the dynamical structure factor, $S(\mathbf{Q},\omega)$~\cite{petsch2023high,scheie2023spin,hase2024inelastic}. However, each sample orientation probes only a limited region of reciprocal space, requiring multiple measurements for comprehensive coverage, and sufficiently large, high-quality single crystals are often difficult to grow. By contrast, powder samples are generally easier to prepare in the quantities required for neutron-scattering experiments. Powder INS measures the orientationally averaged response, $S(|\mathbf{Q}|,\omega)$, sacrificing directional information in exchange for greater experimental accessibility and broader sampling of reciprocal space~\cite{avdoshenko2022spin,su2024uncovering,cheng2023database}. These complementary measurement modalities therefore impose distinct observational constraints on the same underlying Hamiltonian. Moreover, multimodal experiments often differ in their coordinate representations, spatial and energy resolutions, selection rules, and noise characteristics, making it challenging to integrate heterogeneous datasets in a manner that is both mathematically rigorous and physically interpretable.

Artificial intelligence (AI) and machine learning (ML) have been playing increasingly important roles in bridging experimental observations and theories and computations~\cite{chen2021machine,plumley2024ultrafast,horwath2024ai,li2025powder,carleo2017solving,butler2018machine}. Recent advances in simulation-based inference, neural posterior estimation, and multimodal machine learning \cite{papamakarios2016snpe,
lueckmann2021sbi,
baltrusaitis2019multimodal,
liang2024foundations} have provided powerful new tools for accelerating inverse problems and integrating heterogeneous datasets. Many of these approaches, however, rely on learned latent representations or gridded data parameterizations to relate observations across different modalities. 
In practice, discrepancies between the simulated data used for training and experimental measurements, such as instrumental effects, sample imperfections, and other phenomena not captured by the forward model, can degrade the robustness of learned latent representations and complicate their physical interpretation, particularly in high-dimensional settings.
% More fundamentally, for Hamiltonian inference, the natural shared representation is not a learned latent embedding but the parameter space of the microscopic Hamiltonian itself. 
For the Hamiltonian-inference problem considered here, however, a natural shared representation is provided by the parameter space of the microscopic Hamiltonian itself.
An effective framework for multimodal spectroscopy should therefore connect complementary measurements through this physically meaningful parameter space, propagate uncertainty consistently across measurement modalities, and guide experimental design by identifying the measurements that most effectively resolve the remaining ambiguities.

In this work, we develop a framework that asks not only which Hamiltonian best reproduces an INS spectrum, but also what microscopic information the measurements actually constrain. Our central idea is to treat each spectroscopic modality as an observation mapping from the Hamiltonian parameter space to the experimentally accessible response function. 
This mapping induces an observation geometry on the Hamiltonian parameter space, where stiff directions correspond to parameter combinations that strongly influence the measured spectra, while sloppy directions produce nearly indistinguishable responses and are thus only weakly constrained~\cite{gutenkunst2007universally,transtrum2015perspective,amari2016information}.

To make this observation geometry computationally tractable, we construct Hamiltonian-conditioned, differentiable AI surrogate models that accurately emulate multimodal neutron-scattering spectra, leveraging recent advances in implicit neural representations (INRs) for spectroscopy~\cite{sitzmann2020siren,chitturi2023capturing,chen2025implicit}.
Combined with Bayesian inference, these surrogates enable rapid evaluation of spectral-discrepancy landscapes, quantification of modality-dependent parameter-wise identifiability, and propagation of posterior uncertainty directly within the physical Hamiltonian parameter space. 
Rather than fusing powder and single-crystal INS through an abstract learned embedding, our framework connects them through the shared microscopic Hamiltonian that generates both measurements, thereby keeping inference within this physically meaningful parameter space. 
As a result, posterior information inferred from one modality can naturally inform inference in another, enabling sequential Bayesian inference, joint parameter estimation, and uncertainty-aware experimental design~\cite{chaloner1995bayesian,granade2012robust,huan2013simulation,mcmichael2022simplified,huan2024optimal}.

We demonstrate this framework using simulated and experimental INS measurements of the quantum magnet NiPS\textsubscript{3}. The observation geometry reveals modality-dependent stiff and sloppy Hamiltonian directions, powder-informed posterior initialization reshapes the subsequent single-crystal acquisition strategy, and joint inference on experimental powder and single-crystal spectra yields a posterior distribution of Hamiltonians that reproduce both modalities with high spectral agreement. This work establishes a general framework that uses AI surrogates not only to accelerate Hamiltonian inference, but also to efficiently uncover which microscopic interactions are resolved by each experimental modality, quantify the remaining ambiguities, and guide adaptive experimental design toward measurements with higher expected information gain.

\section*{Results}

\subsection*{Multimodal AI surrogate}

\begin{figure*}
    \includegraphics[width=0.85\linewidth]{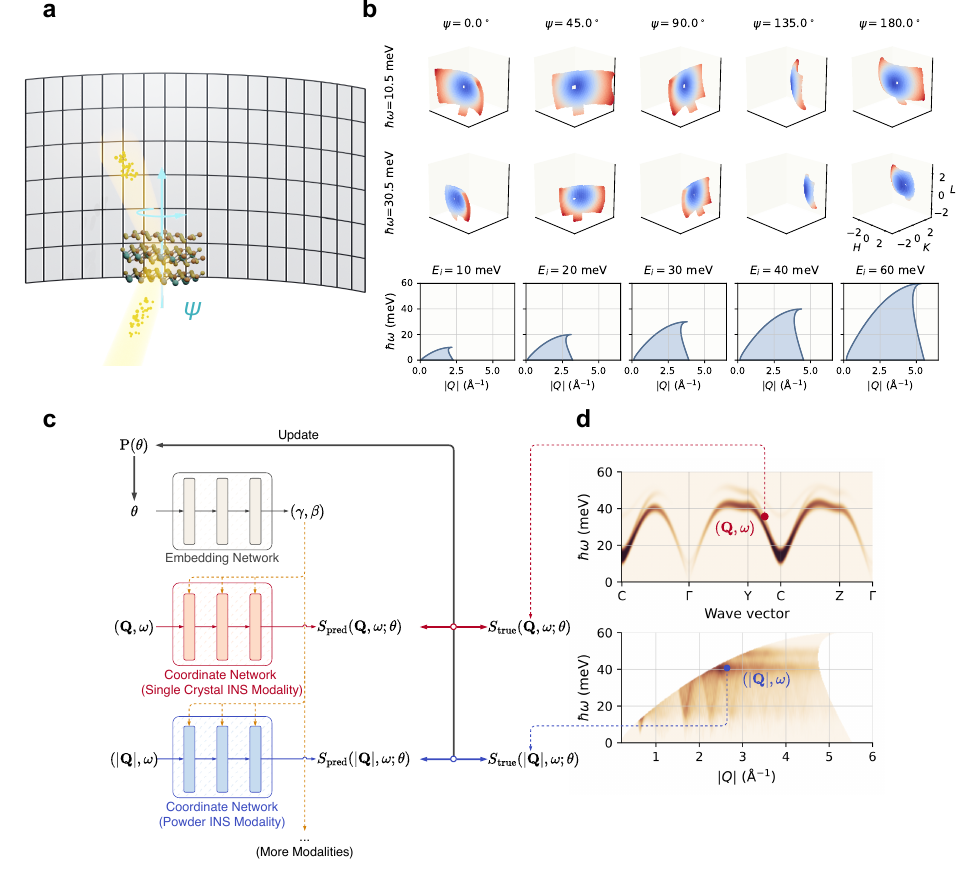}
    \caption{
    \textbf{Multimodal inelastic neutron scattering (INS) measurements and Hamiltonian-conditioned AI surrogate model.}
    \textbf{(a)}
    Incident neutrons scatter from the sample and are recorded by a detector array, sampling a subset of the four-dimensional scattering space $(\mathbf{Q},\omega)$ for single-crystal measurements or a bounded region in $(|\mathbf{Q}|,\omega)$ for powder measurements.
    \textbf{(b)} 
    Experimental control variables and the corresponding measurement coverage. The upper panel illustrates the accessible $(\mathbf{Q},\omega)$ region for different single-crystal sample orientations, while the lower panel shows the accessible $(|\mathbf{Q}|,\omega)$ region for different incident neutron energies in powder measurements.
    \textbf{(c)}
    AI surrogate model based on Hamiltonian-conditioned implicit neural representations (INRs) for multimodal INS. A shared embedding network encodes the Hamiltonian parameters into a common latent representation that conditions modality-specific coordinate networks, which predict scattering intensities at arbitrary $(\mathbf{Q},\omega)$ coordinates for single-crystal measurements and $(|\mathbf{Q}|,\omega)$ coordinates for powder measurements. The trained surrogate enables efficient Bayesian inference of Hamiltonian parameters from multimodal INS measurements.
    \textbf{(d)} 
    Representative single-crystal (top) and powder (bottom) INS spectra illustrating the complementary reciprocal-space coverage of the two measurement modalities.
    }
    \label{fig:workflow_overview}
\end{figure*}

In an INS experiment, illustrated schematically in Fig.~\ref{fig:workflow_overview}(a), an incident neutron beam scatters from the sample, and the scattered neutrons are recorded by an array of detectors as a function of momentum and energy transfer. For a single-crystal sample, the crystal orientation is controlled by a goniometer with rotation angle $\psi$. 
% Each orientation measures a three-dimensional hypersurface of the four-dimensional scattering function, $I \propto S(\mathbf{Q},\omega)$, defined by the detector coverage and the accessible combinations of momentum transfer $\mathbf{Q}=[H,K,L]$ and energy transfer $\omega$.
For a fixed sample orientation $\psi$, the detector array samples the measured intensity $I(\mathbf{Q},\omega)\propto S(\mathbf{Q},\omega)$ over a region of the four-dimensional scattering space determined by the detector coverage and the kinematically accessible combinations of momentum transfer $\mathbf{Q}=[H,K,L]$ and energy transfer $\hbar\omega$.
Representative measurement coverages for different rotation angles are shown in the upper panels of Fig.~\ref{fig:workflow_overview}(b).
For a powder sample, the random orientation of crystallites reduces the measurement to the two-dimensional response $S(Q,\omega)$, where $Q=|\mathbf{Q}|$ is the magnitude of the momentum transfer. The corresponding detector coverage for different incident neutron energies is shown in the lower panels of Fig.~\ref{fig:workflow_overview}(b).

We choose the van der Waals (vdW) quantum magnet NiPS\textsubscript{3} as the representative material system throughout this work. Its spin Hamiltonian is given by~\cite{scheie2023spin}
\begin{equation}
    H=\sum_{i,j}J_{ij}\mathbf{S}_{i}\cdot \mathbf{S}_{j}+\sum_{i}\left[A_{x}(S_{i}^{x})^{2}+A_{z}(S_{i}^{z})^{2}\right],
\end{equation}
where $J_{ij}$ denotes the Heisenberg exchange interactions extending from the first- to fourth-nearest neighbors, parameterized by $[J_{\mathrm{1a}},J_{\mathrm{1b}},J_{\mathrm{2a}},J_{\mathrm{2b}},J_{\mathrm{3a}},J_{\mathrm{3b}},J_{\mathrm{4}}]$, while $A_x$ and $A_z$ are the single-ion anisotropy parameters. Ref.~\citenum{scheie2023spin} distinguished $J_{\mathrm{1a}}$ and $J_{\mathrm{1b}}$ to account for the crystallographic distortion while imposing $J_{\mathrm{2a}}=J_{\mathrm{2b}}$ and $J_{\mathrm{3a}}=J_{\mathrm{3b}}$. Because the same symmetry reduction formally permits the corresponding second- and third-neighbor exchanges to be distinct, we relax these constraints and treat $J_{\mathrm{2a}}$, $J_{\mathrm{2b}}$, $J_{\mathrm{3a}}$, and $J_{\mathrm{3b}}$ as independent parameters. These seven exchange and two anisotropy parameters together define a nine-dimensional (9D) Hamiltonian parameter space, which is denoted as $\bm{\theta}=[A_{x},A_{z},J_{\mathrm{1a}},J_{\mathrm{1b}},J_{\mathrm{2a}},J_{\mathrm{2b}},J_{\mathrm{3a}},J_{\mathrm{3b}},J_{\mathrm{4}}]$.

To enable rapid, differentiable forward evaluations, we construct an AI surrogate model that maps Hamiltonian parameters to INS spectra. To accommodate multiple INS measurement modalities, the model consists of a shared Hamiltonian parameter embedding network coupled with modality-specific INRs \cite{sitzmann2020siren}, which serve as coordinate networks. The embedding network conditions the modality-specific INRs on the Hamiltonian parameters, allowing different parameter sets to generate distinct dynamical structure factors, $S(\mathbf{Q},\omega)$. For example, the coordinate network for single-crystal INS maps the momentum and energy transfer coordinates, $(\mathbf{Q},\omega)$, to the scattering intensity. Combined with the parameter embedding network, the surrogate realizes the conditional mapping $(\mathbf{Q},\omega;\bm{\theta}) \mapsto S(\mathbf{Q},\omega)$, where $\bm{\theta}$ denotes the Hamiltonian parameters. By using a shared parameter embedding network to condition multiple observables, the intrinsically multimodal design provides a physics-based, extensible, and flexible framework. The overall architecture is illustrated in Fig.~\ref{fig:workflow_overview}(c), with additional details on the datasets and model architecture provided in Methods and \ref{sec_SN:ai_surrogate}.

The differentiability and computational efficiency of the surrogate model enable seamless integration with Bayesian inference and Bayesian experimental design \cite{chaloner1995bayesian,granade2012robust,huan2013simulation,mcmichael2022simplified,huan2024optimal}. In particular, the surrogate-predicted spectra, $S_{\mathrm{pred}}$, serve as the forward model in the likelihood function, where likelihoods are evaluated by direct comparison with the measured spectra, $S_{\mathrm{targ}}$, as illustrated in Fig.~\ref{fig:workflow_overview}(d).

\subsection*{Observation-induced geometry of Hamiltonian parameter space}

\begin{figure*}
    \centering
    \includegraphics[width=.85\linewidth]{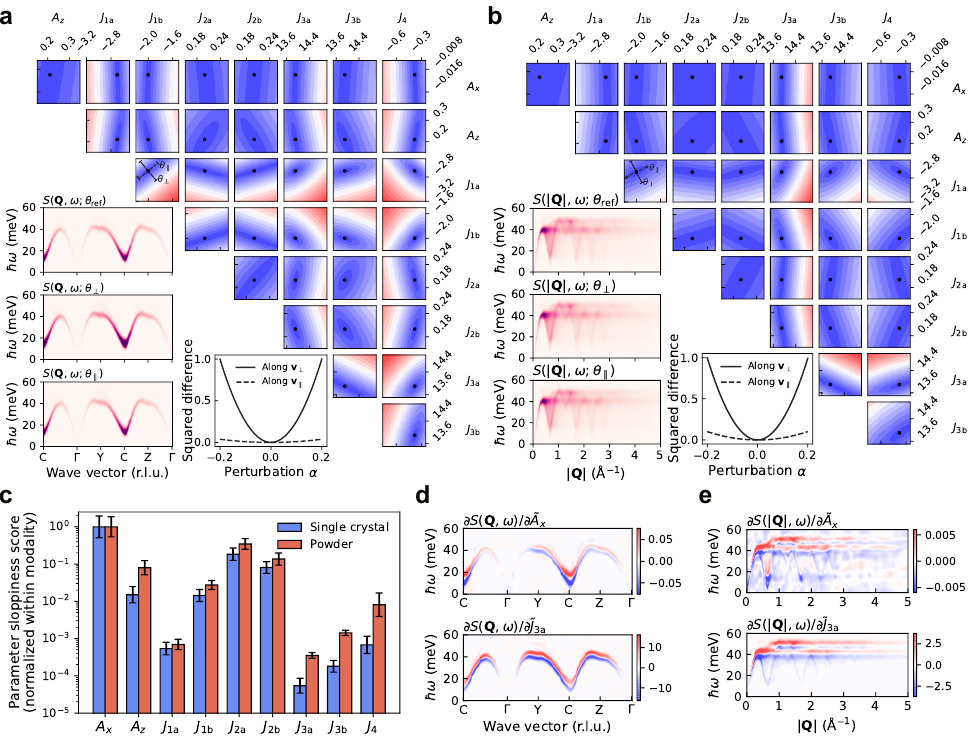}
    \caption{
    \textbf{Observation landscapes and local parameter identifiability across INS modalities.}
    \textbf{a}, Local observation landscape for the single-crystal INS modality, centered at the reference Hamiltonian $\bm{\theta}_{\mathrm{ref}}$. The corner plot shows two-parameter slices of the spectral-discrepancy landscape between the reference spectrum, $S(\mathbf{Q},\omega;\bm{\theta}_{\mathrm{ref}})$, and spectra generated by perturbed Hamiltonians, $S(\mathbf{Q},\omega;\bm{\theta})$. Circular markers indicate $\bm{\theta}_{\mathrm{ref}}$. In the $(J_{\mathrm{1a}},J_{\mathrm{1b}})$ subspace, the solid and dashed lines denote perturbations along the eigenvectors, $\mathbf{v}_{\perp}$ and $\mathbf{v}_{\parallel}$, of the local observation metric tensor $\mathbf{G}$, with eigenvalues satisfying $\lambda_{\perp}\geq\lambda_{\parallel}$. These directions correspond to the local stiff and sloppy directions in Hamiltonian parameter space, respectively. Representative spectra and the corresponding normalized squared differences demonstrate that perturbations of comparable magnitude along $\mathbf{v}_{\perp}$ produce substantially larger spectral changes than perturbations along $\mathbf{v}_{\parallel}$.
    \textbf{b}, Corresponding local observation landscape for the powder INS modality, illustrating the distinct observation geometry induced by orientational averaging.
    \textbf{c}, Parameter-wise sloppiness scores for single-crystal and powder INS, computed from the diagonal elements of the inverse local metric tensor, $\tilde{\mathbf{G}}^{-1}$, and normalized independently within each modality such that the maximum sloppiness score equals to $1$. Larger values indicate Hamiltonian parameters that are less well constrained by the corresponding measurement modality. 
    \textbf{d}, Representative partial derivatives of the dynamical structure factor with respect to selected normalized Hamiltonian parameters for the single-crystal modality, $\partial S(\mathbf{Q},\omega)/\partial \tilde{A}_x$ and $\partial S(\mathbf{Q},\omega)/\partial \tilde{J}_{3a}$.
    \textbf{e}, Representative partial derivatives of the dynamical structure factor with respect to selected normalized Hamiltonian parameters for the powder modality, $\partial S(|\mathbf{Q}|,\omega)/\partial \tilde{A}_x$ and $\partial S(|\mathbf{Q}|,\omega)/\partial \tilde{J}_{3a}$.
    }
    \label{fig:landscape}
\end{figure*}

We define the empirical spectral distance between two Hamiltonian
parameter sets using the normalized $L^2$ distance between their
corresponding scalar scattering intensities,
\begin{equation}
\label{eq:spectral_distance}
d_N(\bm{\theta},\bm{\theta}')
=
\left[
\frac{1}{N}
\sum_{i=1}^{N}
\left|
S(x_i;\bm{\theta})-S(x_i;\bm{\theta}')
\right|^2
\right]^{1/2},
\end{equation}
where $x_i=(\mathbf{Q}_i,\omega_i)$ are the sampled
momentum--energy coordinates. This is a metric on the sampled
spectral vectors and induces a pseudometric on the Hamiltonian
parameter space; it becomes a metric there if the forward map is
injective. To characterize its local geometry, we define the smooth
squared-distance landscape
$f(\bm{\theta})=d_N^2(\bm{\theta},\bm{\theta}_{\mathrm{ref}})$, where the reference Hamiltonian $\bm{\theta}_{\mathrm{ref}}$ is taken from Ref.~\citenum{scheie2023spin}, whose parameter values are listed in \ref{sec_SN:hamiltonian_param_vals}.

An example of the local spectral-discrepancy landscape, $f(\bm{\theta})$, for single-crystal INS is shown in Fig.~\ref{fig:landscape}(a). Several notable features emerge. First, the landscape is relatively flat along the $A_x$ and $A_z$ directions, indicating that these parameters weakly influence $S(\mathbf{Q},\omega)$ near the reference Hamiltonian. Second, correlated parameter combinations involving $(J_{\mathrm{1a}},J_{\mathrm{1b}})$ and $(J_{\mathrm{2a}},J_{\mathrm{2b}})$ produce similar spectra, leading to elongated valleys in the observation landscape. In contrast, anti-correlated variations involving $J_{\mathrm{3a}}$ and $J_{\mathrm{3b}}$ produce only small spectral changes. 
These observations illustrate the central principle: spectral distinguishability is governed by the geometry induced by the forward mapping, $\bm{\theta}\mapsto S(\mathbf{Q},\omega;\bm{\theta})$, rather than by Euclidean distance in the Hamiltonian parameter space.

To characterize the observation-induced geometry more generally, 
% let $x=(\mathbf{Q},\omega)\in\Omega$ denote a point in momentum--energy space, 
let $x\in\Omega$ denote a coordinate point in the observation space,
where $\Omega$ is the experimentally accessible measurement domain. The spectroscopic forward map $S(x;\bm{\theta})$ induces a local metric on the Hamiltonian parameter space through the normalized $L^2(\Omega)$ inner product,
\begin{widetext}
\begin{equation}
\label{eqn:observation_metric}
G_{mn}(\bm{\theta})=
\left\langle\partial_{\theta_m} S(x\,;\,\bm{\theta}),\partial_{\theta_n} S(x\,;\,\bm{\theta})\right\rangle_{\Omega}=
\frac{1}{V_\Omega}\int_{\Omega}\partial_{\theta_m} S(x;\bm{\theta})\,\partial_{\theta_n} S(x;\bm{\theta})\,\mathrm{d}x,
\end{equation}
\end{widetext}
where $V_\Omega$ denotes the measure of the accessible momentum--energy domain and $\theta_{m}$ denotes the $m$th component of $\bm{\theta}$.
For INS, $x=(\mathbf{Q},\omega)$, and the normalized inner product reads
\begin{equation}
\label{eqn:observation_metric}
G_{mn}(\bm{\theta})=\frac{1}{V_{\Omega}}\int_{\Omega}\frac{\partial S(\mathbf{Q},\omega;\bm{\theta})}{\partial \theta_m}\frac{\partial S(\mathbf{Q},\omega;\bm{\theta})}{\partial \theta_n}\,\mathrm{d}^{3}\mathbf{Q}\,\mathrm{d}\omega,
\end{equation}
where $V_{\Omega}=\int_{\Omega} \mathrm{d}^{3}\mathbf{Q}\,\mathrm{d}\omega$ is the accessible measurement volume. 
In essence, this defines the intensity-induced $L^2$ pullback metric.
In practice, the integral is approximated using Monte Carlo sampling,
\begin{equation}
\label{eqn:observation_metric_discrete}
G_{mn}(\bm{\theta}) \approx \frac{1}{N} \sum_{i=1}^{N} \frac{\partial S(x_i;\bm{\theta})}{\partial \theta_m} \frac{\partial S(x_i;\bm{\theta})}{\partial \theta_n},
\end{equation}
where $x_i=(\mathbf{Q}_i,\omega_i)$ are sampled uniformly over $\Omega$ with respect to the normalized momentum-energy volume measure. Unless otherwise specified, we use $N=25{,}000$ samples to estimate the observation metric.

The eigenstructure of the pullback Gram matrix characterizes the local spectral sensitivity in the chosen Hamiltonian parameterization. Eigenvectors corresponding to large eigenvalues define \emph{stiff} directions, along which small parameter perturbations produce substantial changes in the predicted spectra. Eigenvectors with small eigenvalues define \emph{sloppy} directions, along which local parameter perturbations produce comparatively small spectral changes. These directions characterize sensitivity under the selected $L^2$ measure and do not by themselves incorporate counting noise, acquisition time, or prior information.

An example within the $(J_{\mathrm{1a}},J_{\mathrm{1b}})$ subspace is shown in the corner plot of Fig.~\ref{fig:landscape}(a). The perturbed Hamiltonians $\bm{\theta}_{\perp}$ and $\bm{\theta}_{\parallel}$ are displaced by an equal distance of $0.2~\mathrm{meV}$ from the reference Hamiltonian $\bm{\theta}_{\mathrm{ref}}$ along the stiff and sloppy eigen-directions of the local 2D subspace metric tensor $\mathbf{G}(\bm{\theta}_{\mathrm{ref}})$, respectively. The corresponding spectra, $S(\mathbf{Q},\omega;\bm{\theta}_{\perp})$ and $S(\mathbf{Q},\omega;\bm{\theta}_{\parallel})$, are shown in the lower-left panels of Fig.~\ref{fig:landscape}, while the corresponding 9D eigenvectors are visualized in \ref{sec_SN:hidim_eigen}. Perturbations along the stiff direction produce substantially larger spectral changes relative to $S(\mathbf{Q},\omega;\bm{\theta}_{\mathrm{ref}})$ than perturbations along the sloppy direction. The normalized squared spectral differences, shown in the lower-right panels of Fig.~\ref{fig:landscape}, further highlight the different sensitivities along these two directions.
These results demonstrate that parameter-wise resolving power is fundamentally governed by the observation geometry induced by the forward map. Rather than merely identifying which Hamiltonian best reproduces the measurements, the observation geometry characterizes what information the measurements can and cannot resolve.

In Fig.~\ref{fig:landscape}(b), we present the corresponding local observation landscape for the powder INS modality. 
Relative to single-crystal INS, powder averaging produces broader low-discrepancy regions in Hamiltonian parameter space, reflecting the loss of directional information. As a result, orientationally averaged spectra are less effective at distinguishing correlated parameter variations.
The representative powder spectra and their corresponding squared-difference maps again reveal stiff and sloppy parameter combinations, although the spectral response is generally less discriminative than in single-crystal INS.
Despite this reduced parameter-wise resolving power, the observation geometry demonstrates that powder INS remains informative for Hamiltonian inference. In particular, it provides meaningful constraints on the Hamiltonian parameter space, making it a practical alternative when sufficiently large single crystals are unavailable or when rapid measurements with limited beam time are preferred.
The complete 9D eigen-structures of the observation metric for both single-crystal and powder modalities are provided in \ref{sec_SN:hidim_eigen}.

To compare the local identifiability of Hamiltonian parameters with different scales, we evaluate the observation metric in locally normalized coordinates.
For a metric evaluated at a reference parameter point $\hat{\bm{\theta}}$, we define the dimensionless coordinates
\begin{equation}
\label{eq:relative_parameters}
\tilde{\theta}_m = \frac{\theta_m-\hat{\theta}_{m}}{|\hat{\theta}_{m}|}.
\end{equation}
The corresponding scaled observation metric is
\begin{equation}
\label{eq:relative_scaled_metric}
\tilde{\mathbf{G}}(\hat{\bm{\theta}}) = \mathbf{D}(\hat{\bm{\theta}}) \mathbf{G}(\hat{\bm{\theta}}) \mathbf{D}(\hat{\bm{\theta}}), \quad \mathbf{D}(\hat{\bm{\theta}}) = \mathrm{diag}\!\left( |\hat{\theta}_{1}|, \ldots, |\hat{\theta}_{M}| \right),
\end{equation}
where $M$ is the number of Hamiltonian parameters.
This scaled metric removes the dependence of the local geometry on the absolute magnitudes of the Hamiltonian parameters, enabling more intuitive comparisons of parameter identifiability across different Hamiltonian parameters.
We quantify the sloppiness of parameter $\theta_m$ using the corresponding diagonal element of the inverse scaled metric,
\begin{equation}
\label{eq:parameter_sloppiness_score}
\mathcal{S}_m(\hat{\bm{\theta}}) = \left[ \left( \tilde{\mathbf{G}}(\hat{\bm{\theta}}) + \epsilon \mathbf{I} \right)^{-1} \right]_{mm},
\end{equation}
where $\epsilon$ is a small numerical regularizer ($\epsilon=10^{-12}$ throughout this work). Larger values of $\mathcal{S}_m$ indicate that parameter $\theta_m$ is less locally identifiable from the measured spectra. 

To obtain a representative measure of parameter identifiability across the Hamiltonian parameter space, we evaluate the sloppiness scores at 100 randomly sampled Hamiltonians $\hat{\bm{\theta}}$ and report their averages in Fig.~\ref{fig:landscape}(c).
For visualization, the sloppiness scores are normalized within each measurement modality by their largest value. 
% For both powder and single-crystal INS, the transverse anisotropy $A_x$ is identified as the least constrained parameter. 
% Aside from $A_x$, all remaining Hamiltonian parameters exhibit systematically larger normalized sloppiness scores under powder INS than under single-crystal INS (Fig.~\ref{fig:landscape}(c)). 
For both powder and single-crystal INS, the transverse anisotropy $A_{x}$ is identified as the least constrained parameter and therefore sets the normalization scale. 
Because the scores are normalized independently, this comparison concerns relative parameter-wise sloppiness within each modality rather than the absolute magnitude of the metric between modalities.
Relative to this common modality-wise reference, all remaining Hamiltonian parameters exhibit systematically larger normalized sloppiness scores under powder INS than under single-crystal INS (Fig.~\ref{fig:landscape}(c)). This indicates that the parameter-wise resolving power is less differentiated in the powder modality, with a broader set of Hamiltonian parameters remaining relatively sloppy compared with the least constrained direction.
This trend reflects the orientational averaging inherent in powder measurements, which suppresses directional spectral features and expands the set of Hamiltonian perturbations that produce nearly indistinguishable spectra.

The partial derivatives shown in Fig.~\ref{fig:landscape}(d,e) provide an intuitive visualization of the local observation geometry by revealing how infinitesimal perturbations of individual Hamiltonian parameters modify the measured spectra. Consistent with the sloppiness analysis, parameters with weaker spectral sensitivity generally exhibit smaller derivative magnitudes over the measured momentum--energy domain. Because these derivatives are obtained directly by automatic differentiation of the AI surrogate, they can be evaluated efficiently, in contrast to conventional forward solvers where derivative information is typically unavailable or substantially more expensive to compute numerically.

More generally, the parameter-wise sloppiness profile provides a quantitative description of how effectively a given measurement modality constrains different directions in the Hamiltonian parameter space. This observation suggests a practical principle for multimodal experimental design: complementary measurements are most informative when they exhibit complementary sloppy directions, allowing one modality to constrain parameter combinations that remain weakly identifiable in another. The powder--single-crystal combination considered here serves primarily as a proof of concept, since powder INS is effectively an orientationally averaged projection of single-crystal scattering. Even greater gains can be expected when combining experimental probes with genuinely complementary observation operators and correspondingly distinct observation geometries and sloppiness profiles.

\subsection*{Posterior-guided multimodal Hamiltonian inference and adaptive experimental design}

\begin{figure*}
    \centering
    \includegraphics[width=0.85\linewidth]{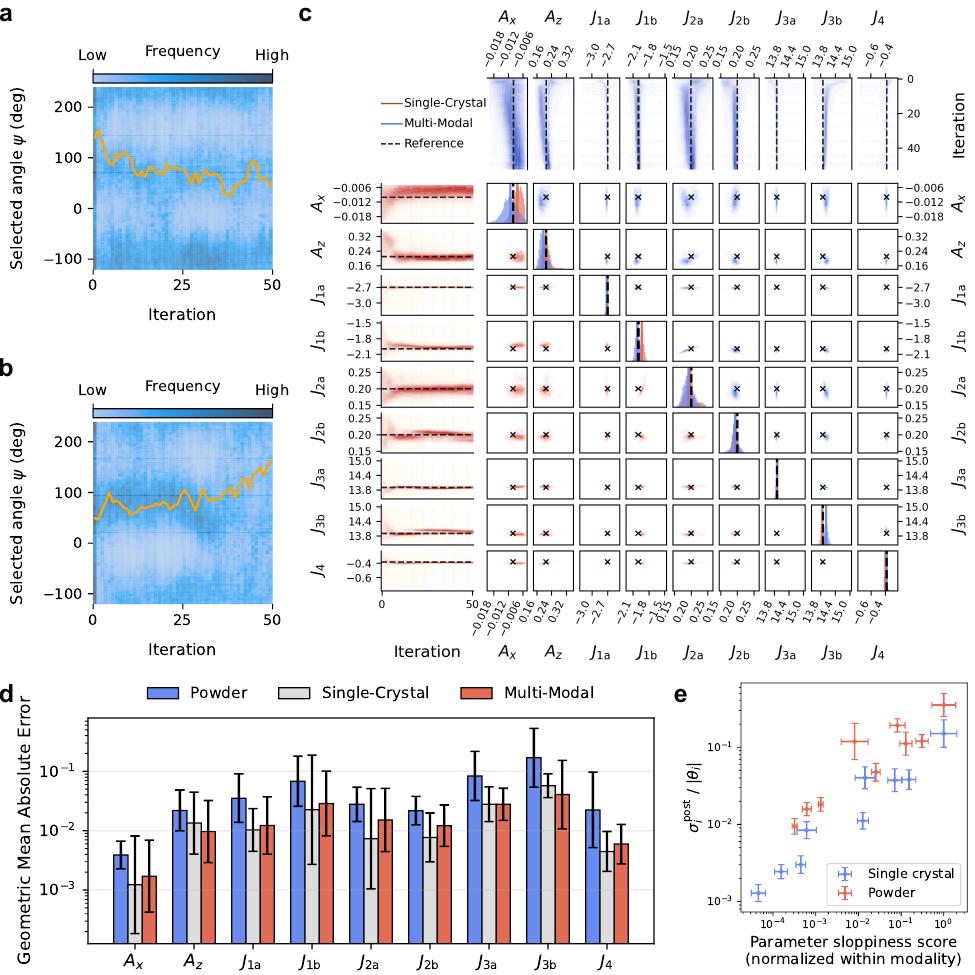}
    \caption{
    \textbf{Bayesian experimental design and multimodal Hamiltonian inference enabled by AI surrogates.}
    \textbf{a}, 
    Adaptive experimental design on orientation $\psi$ for single-crystal INS without powder-derived prior information. The orange curve shows the sequence of selected sample orientations $\psi$ for a representative simulated experiment generated from $S(\mathbf{Q},\omega;\bm{\theta}_{\mathrm{ref}})$ with Poisson noise. 
    % The heatmap shows the distribution of suggested orientations across Hamiltonian parameter vectors $\bm{\theta}$ sampled from the initial uniform prior $P_{0}$.
    The heatmap shows the average predictive-variance utility over $10$ simulated experiments with ground-truth Hamiltonians sampled from the initial prior.
    \textbf{b}, 
    Adaptive experimental design on orientation $\psi$ for single-crystal INS initialized with a posterior obtained from powder INS measurements. Compared with panel (a), the suggested orientations shift toward different regions, reflecting the additional information contributed by the powder modality.
    \textbf{c}, 
    Posterior distributions of Hamiltonian parameters. The top and left panels show the marginal posterior evolution as a function of measurement steps, while the corner plot illustrates the posterior distribution after ten single-crystal measurements.
    \textbf{d}, 
    Parameter estimation error for powder-only, single-crystal-only, and multimodal inference. Bars show the geometric mean of absolute error across benchmark realizations over sampled $\bm{\theta}$, computed as $y=\exp[\langle \log |\hat{\theta}-\theta^{\ast}| \rangle]$ and displayed on a logarithmic scale. Error bars indicate one log-standard-deviation interval, where $\sigma_{\log}=\mathrm{std}[\log |\hat{\theta}-\theta^\ast|]$, corresponding to multiplicative intervals $[y\exp(-\sigma_{\log}),\,y\exp(\sigma_{\log})]$.
    \textbf{e}, 
    Relationship between the parameter sloppiness score and the residual posterior uncertainty after sequential single-crystal inference. Each point corresponds to one Hamiltonian parameter. The positive correlation demonstrates that the local observation geometry provides a useful predictor of the uncertainty remaining after Bayesian inference.
    }
    \label{fig:siml_benchmark}
\end{figure*}

The proposed framework advances our understanding of how Hamiltonian parameters are jointly constrained by multimodal experimental observables. By learning a differentiable and computationally efficient mapping between the Hamiltonian parameter space and INS spectra, the AI surrogate provides the foundation for Bayesian experimental design, enabling posterior-guided Hamiltonian inference that integrates heterogeneous measurements and adaptively designs subsequent experiments.

In INS, single-crystal measurements preserve directional momentum information and can resolve fine features of the dynamical structure factor, $S(\mathbf{Q},\omega)$. However, each crystal orientation probes only a subset of reciprocal space, making comprehensive reciprocal-space coverage experimentally demanding. In contrast, powder INS rapidly provides an orientationally averaged overview of the excitation spectrum, although the averaging process obscures directional information.
These complementary measurements therefore induce distinct observation geometries for the same underlying Hamiltonian: powder measurements provide an efficient global characterization, while single-crystal measurements subsequently resolve the remaining microscopic ambiguities.

To exploit this complementarity, we perform sequential Bayesian inference in which information inferred from powder INS is used to initialize the subsequent single-crystal analysis. We first infer the powder posterior $P_{\mathrm{pd}}(\bm{\theta})$, represented computationally by the particle approximation $\mathcal{P}_{\mathrm{pd}}$. The initial particle distribution for the single-crystal stage is then constructed from $\mathcal{P}_{\mathrm{pd}}$ using the partial posterior initialization procedure described in Methods. After acquiring the $(t+1)$th single-crystal measurement at orientation $\psi^{(t+1)}$, the posterior is updated according to
\begin{equation} \label{eq:multimodal_sequential_bayes}
P_{\mathrm{sc}}^{(t+1)}(\bm{\theta})
\propto
P\!\left(
S_{\mathrm{sc}}^{(t+1)}
\mid
\bm{\theta},
\psi^{(t+1)}
\right)
P_{\mathrm{sc}}^{(t)}(\bm{\theta}),
\end{equation}
where $S_{\mathrm{sc}}^{(t+1)}$ denotes the measured single-crystal spectrum acquired at iteration $t+1$, and $\psi^{(t+1)}$ is the corresponding sample orientation.
% The initial prior for the single-crystal stage is therefore given by the posterior inferred from the powder measurements,
% \begin{equation}
% \label{eq:powder_initialized_prior}
% P_{\mathrm{sc}}^{(0)}(\bm{\theta}) = P_{\mathrm{pd}}(\bm{\theta}),
% \end{equation}
% where $P_{\mathrm{pd}}(\bm{\theta})$ is the posterior inferred from the powder INS data.

Unlike many multimodal learning approaches~\cite{baltrusaitis2019multimodal,liang2024foundations}, which fuse heterogeneous observations through learned latent representations, our framework connects powder and single-crystal measurements directly through the shared Hamiltonian parameter space $\bm{\theta}$.
The modality-specific surrogate branches map the same Hamiltonian parameters to powder and single-crystal response functions in their respective observation spaces.
Consequently, posterior uncertainty inferred from powder INS can be propagated directly through the shared Hamiltonian parameterization to predict single-crystal spectra, enabling principled experimental planning in the subsequent measurement stage.

The next single-crystal measurement is selected using Bayesian experimental design. Here, we demonstrate the experimental design capability using the crystal orientation $\psi$ while this can be generalized to other experimental settings. We approximate the expected information gain using the predictive spectral variance~\cite{huan2013simulation,mcmichael2022simplified,chen2023bayesian,chen2025implicit}, leading to the utility function
\begin{equation}
\label{eq:bed_utility}
U^{(t+1)}(\psi)=\frac{1}{N_{\mathbf{Q},\omega}^{(\psi)}}\sum_{(\mathbf{Q},\omega)\in \Omega_{\psi}}\mathrm{Std}_{\bm{\theta}\sim P_{\mathrm{sc}}^{(t)}(\bm{\theta})}\left[S_{\mathrm{sc}}(\mathbf{Q},\omega;\bm{\theta})\right],
\end{equation}
where $\Omega_{\psi}$ denotes the momentum--energy region accessible at experimental setting $\psi$, and $N_{\mathbf{Q},\omega}^{(\psi)}$ is the number of sampled $(\mathbf{Q},\omega)$ points within this region. In practice, the utility is further adjusted to account for experimental preferences, including penalties for repeated measurements and large changes in sample orientation (see Methods and Supplementary Information). The next experimental setting (i.e., orientation $\psi^{(t+1)}$) is then selected by maximizing $U^{(t+1)}(\psi)$ over the candidate orientations.

We demonstrate the proposed workflow using simulated INS experiments (Fig.~\ref{fig:siml_benchmark}). For each ground-truth Hamiltonian used in the simulated experiments, synthetic single-crystal spectra are generated from $S_{\mathrm{sc}}(\mathbf{Q},\omega;\bm{\theta})$ with Poisson counting noise and treated as experimental observations.
Figures~\ref{fig:siml_benchmark}(a) and (b) illustrate the adaptive single-crystal acquisition strategies obtained without and with powder-informed initialization, respectively. 
In each panel, the heatmap shows the average predictive-variance utility, $U^{(t)}(\psi)$, computed from $10$ simulated experiments whose ground-truth Hamiltonians are randomly sampled from the initial uniform prior $P^{(0)}(\bm{\theta})$ defined in Methods. 

The powder-informed initialization reshapes the posterior uncertainty before the single-crystal stage begins, altering the predictive utility landscape over the candidate crystal orientations. Consequently, the preferred measurement sequence adapts to the remaining uncertainty, leading to different orientation trajectories than those obtained from an uninformative prior.
The overlaid trajectory shows the sequence of crystal orientations selected for a representative simulated experiment generated from the same reference Hamiltonian, $\bm{\theta}_{\mathrm{ref}}$, used in Fig.~\ref{fig:landscape}. Figure~\ref{fig:siml_benchmark}(c) compares the posterior evolution for the single-crystal-only and powder-initialized protocols. The marginal distributions along the top and left illustrate the progressive reduction of parameter uncertainty with successive measurements, while the lower-right corner plot shows the final posterior distribution after ten single-crystal acquisitions.

Figure~\ref{fig:siml_benchmark}(d) compares the Hamiltonian parameter estimation errors obtained using powder-only, single-crystal-only, and staged multimodal acquisition protocols. Consistent with the parameter-wise sloppiness analysis in Fig.~\ref{fig:landscape}(c), powder-only inference produces the largest errors for most parameters because orientational averaging removes directional information that is retained in single-crystal measurements. Consequently, single-crystal INS generally achieves the lowest point-estimation errors, reflecting its ability to resolve the full directional structure of $S(\mathbf{Q},\omega)$.

Although powder-initialized measurement does not uniformly reduce the point-estimation error relative to the single-crystal-only measurement, it produces narrower posterior distributions for sloppy Hamiltonian parameters including $A_{x}$, $J_{\mathrm{2a}}$, and $J_{\mathrm{2b}}$ (Fig.~\ref{fig:siml_benchmark}(c)) compared with single-crystal-only measurements. This result indicates that powder measurements effectively concentrate the posterior distribution before single-crystal measurements are acquired, allowing the subsequent measurements to focus on the remaining parameter ambiguities. Rather than providing fundamentally independent information beyond an ideal, fully sampled single-crystal experiment, powder INS serves as an experimentally efficient initialization that enables the subsequent single-crystal measurement budget to be allocated more effectively.

To assess whether the local observation geometry predicts the practical identifiability of Hamiltonian parameters, Fig.~\ref{fig:siml_benchmark}(e) compares the parameter sloppiness scores with the normalized posterior uncertainties obtained after Bayesian inference. The horizontal axis shows the sloppiness score computed from the inverse scaled observation metric, while the vertical axis shows the posterior standard deviation normalized by the magnitude of the corresponding reference parameter, $\sigma_i^{\mathrm{post}}/|\theta_i^{\mathrm{ref}}|$. The positive correlation demonstrates that the local observation geometry provides an effective \textit{a priori} predictor of the parameter uncertainty remaining after inference.
This result suggests that the observation geometry can serve as a diagnostic of the parameter-resolving capability of a measurement configuration solely from the forward model, before experimental data are acquired.

The powder--single-crystal example presented here serves primarily as a proof of concept for the proposed multimodal framework. More generally, the same methodology applies to any combination of experimental probes whose observations can be linked through a shared microscopic Hamiltonian. 
% In this setting, the common Hamiltonian parameter space provides a physically interpretable parameterization through which posterior information can be propagated across measurement modalities, enabling principled sequential inference and adaptive experimental design.

\subsection*{Validation with experimental multimodal INS data}

\begin{figure*}
    \includegraphics[width=0.85\linewidth]{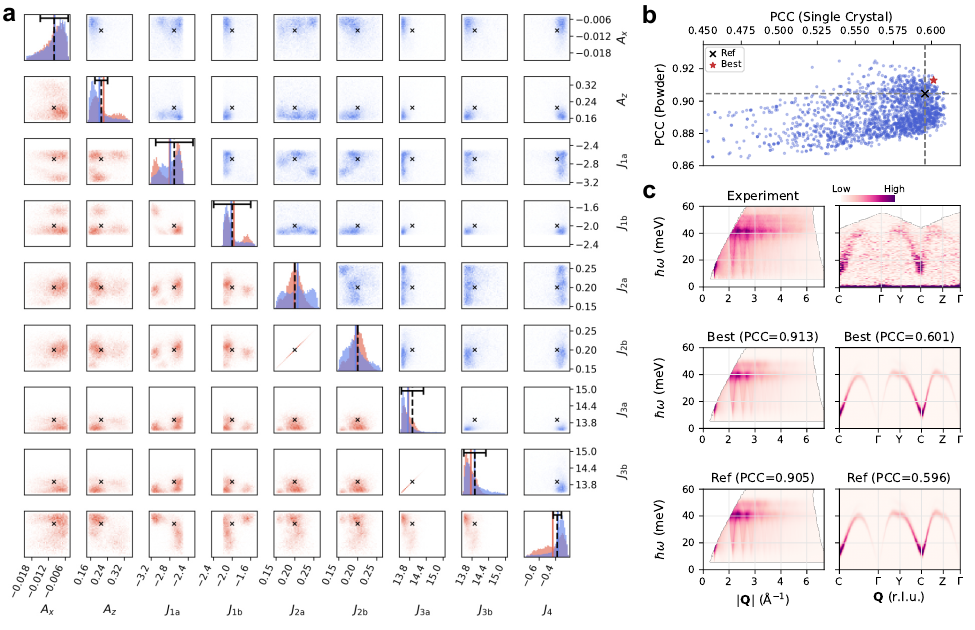}
    \caption{
    \textbf{Validation with experimental multimodal INS data.}
    \textbf{a}, Posterior distributions inferred from multimodal INS measurements with (red) and without (blue) the exchange-parameter constraints $J_{\mathrm{2a}}=J_{\mathrm{2b}}$ and $J_{\mathrm{3a}}=J_{\mathrm{3b}}$. The corner plot shows the final posterior distributions and parameter correlations, while the diagonal panels show the corresponding marginal posterior distributions. Black crosses and dashed vertical lines denote the reference Hamiltonian parameters $\bm{\theta}_{\mathrm{ref}}$, and horizontal error bars indicate the corresponding one-standard-deviation uncertainties reported in Ref.~\citenum{scheie2023spin}.
    \textbf{b}, Pearson correlation coefficients (PCCs) for powder and single-crystal INS evaluated using $2{,}048$ samples drawn from the unconstrained posterior. The selected ``best'' Hamiltonian $\bm{\theta}_{\mathrm{best}}$ is the posterior sample with the smallest normalized distance to the ideal agreement point, as defined in Eq.~\eqref{eq:theta_best}.
    \textbf{c}, Comparison of powder (left) and single-crystal (right) INS spectra. Within each column, from top to bottom: experimental spectrum, prediction from the selected posterior sample $\bm{\theta}_{\mathrm{best}}$, and prediction from the reference Hamiltonian $\bm{\theta}_{\mathrm{ref}}$.
    }
    \label{fig:expt_conerplot_spectrograms}
\end{figure*}

We now apply our multimodal AI surrogate to infer Hamiltonian parameters from experimental INS data collected from powder and single-crystal NiPS\textsubscript{3} samples. Details about experimental data preparation are presented in Methods. As in the staged workflow introduced in the previous section, the two measurement modalities are coupled through the shared Hamiltonian parameter space. 
The objective is therefore not simply to fit each spectrum independently, but to infer a posterior distribution over Hamiltonian parameters that is simultaneously consistent with both powder and single-crystal observations.

Figure~\ref{fig:expt_conerplot_spectrograms}(a) shows the posterior distributions obtained from the two-stage Bayesian inference using the multimodal surrogate. The blue posterior corresponds to the unconstrained nine-parameter Hamiltonian, whereas the red posterior incorporates the symmetry constraints $J_{\mathrm{2a}}=J_{\mathrm{2b}}$ and $J_{\mathrm{3a}}=J_{\mathrm{3b}}$ following Ref.~\citenum{scheie2023spin}, thereby reducing the parameter space from nine to seven dimensions. The constrained posterior is centered near the literature-reported Hamiltonian, while the unconstrained model reveals the associated uncertainty and parameter correlations when these symmetry constraints are relaxed.

Despite the additional degrees of freedom in the unconstrained model, the marginal posteriors of the remaining parameters are broadly consistent between the two formulations. The posterior widths and correlations further indicate that some exchange parameters remain more weakly constrained than others, such as $A_{x}$, $A_{z}$, $J_{\mathrm{2a}}$, and $J_{\mathrm{2b}}$, consistent with the sloppy directions identified by the local observation geometry. 

This extension from a seven-dimensional to a nine-dimensional Hamiltonian is computationally nontrivial for conventional grid-based forward-model evaluation. For a fixed resolution of $N$ points along each parameter dimension, the number of required forward calculations increases by a factor of $N^2$. 
To put this into perspective, using the same computational budget as the surrogate training dataset used in this work ($18{,}000$ forward-model evaluations; see Methods), a direct grid search over the nine-dimensional Hamiltonian space would provide only about three grid points along each parameter dimension, making brute-force Hamiltonian fitting impractical. Instead, these forward-model evaluations are used to train a differentiable neural surrogate that can subsequently be evaluated rapidly throughout the Hamiltonian parameter space, enabling efficient Bayesian inference and experimental design.

Having obtained the unconstrained posterior, we next assess how well its Hamiltonian samples jointly reproduce the experimental powder and single-crystal measurements. We draw $2{,}048$ samples from the posterior and use independent Sunny calculations at the corresponding Hamiltonian parameter sets to generate the associated spectra for comparison with experiment. The scatter plot in Fig.~\ref{fig:expt_conerplot_spectrograms}(b) compares the Pearson correlation coefficients (PCCs) between the experimental data and the corresponding Sunny-generated LSWT spectra for each sampled Hamiltonian. The horizontal and vertical axes denote $\mathrm{PCC}_{\mathrm{sc}}$ (PCC against the single-crystal experimental data) and $\mathrm{PCC}_{\mathrm{pd}}$ (PCC against the powder experimental data), respectively, so that each point represents a single Hamiltonian evaluated simultaneously against both experimental modalities. Larger PCC values therefore indicate better joint agreement with the measured spectra.
A pixel-wise decomposition illustrating the contributions to the PCC metric is provided in \ref{sec_SN:pixelwise_pcc}.

The scatter plot is overlaid with the reference Hamiltonian $\bm{\theta}_{\mathrm{ref}}$ reported in Ref.~\citenum{scheie2023spin}, which was obtained from single-crystal measurements. Although individual posterior samples can fit one modality better than the other, the joint posterior-predictive comparison identifies Hamiltonians that achieve balanced agreement across both measurements.

In particular, let $\mathcal{E}$ denote the set of posterior samples whose single-crystal and powder PCCs are no smaller than the corresponding reference values:
\begin{align*}
    \mathcal{E} = \big\{ \bm{\theta}^{(k)}:
    & \mathrm{PCC}_{\mathrm{sc}}(\bm{\theta}^{(k)}) \geq \mathrm{PCC}_{\mathrm{sc}}(\bm{\theta}_{\mathrm{ref}}),\\
    & \mathrm{PCC}_{\mathrm{pd}}(\bm{\theta}^{(k)}) \geq \mathrm{PCC}_{\mathrm{pd}}(\bm{\theta}_{\mathrm{ref}}) \big\}.
\end{align*}
Assuming that $\mathcal{E}$ is nonempty and that the PCC range is
nonzero for each modality, define
\begin{equation}
\widetilde{\mathrm{PCC}}_{\alpha}(\bm{\theta})
=
\frac{
\mathrm{PCC}_{\alpha}(\bm{\theta})-\min\limits_{\bm{\vartheta}\in\mathcal{E}}\mathrm{PCC}_{\alpha}(\bm{\vartheta})
}{
\max\limits_{\bm{\vartheta}\in\mathcal{E}}\mathrm{PCC}_{\alpha}(\bm{\vartheta})-\min\limits_{\bm{\vartheta}\in\mathcal{E}}\mathrm{PCC}_{\alpha}(\bm{\vartheta})
},
\end{equation}
where $\alpha\in\{\mathrm{sc},\mathrm{pd}\}$. We select the Hamiltonian closest to the component-wise ideal point
$(1,1)$ in the normalized PCC plane:
\begin{equation}
\label{eq:theta_best}
\bm{\theta}_{\mathrm{best}}\in\operatorname*{arg\,min}_{\bm{\theta}\in\mathcal{E}}\sqrt{\left[1-\widetilde{\mathrm{PCC}}_{\mathrm{sc}}(\bm{\theta})\right]^2+\left[1-\widetilde{\mathrm{PCC}}_{\mathrm{pd}}(\bm{\theta})\right]^2}.
\end{equation}

\iffalse
Among posterior samples whose single-crystal and powder PCCs are no worse than the corresponding reference values, we define the ``best'' Hamiltonian, $\bm{\theta}_{\mathrm{best}}$, as the sample
\begin{equation}\label{eq:theta_best}
\bm{\theta}_{\mathrm{best}}=\arg\min_{\bm{\theta}}\sqrt{\left(1-\widetilde{\mathrm{PCC}}_{\mathrm{sc}}\right)^2+\left(1-\widetilde{\mathrm{PCC}}_{\mathrm{pd}}\right)^2},
\end{equation}
that is closest to the ideal agreement point $(\widetilde{\mathrm{PCC}}_{\mathrm{sc}}=1,\widetilde{\mathrm{PCC}}_{\mathrm{pd}}=1)$, where $\widetilde{\mathrm{PCC}}_{\mathrm{sc}}$ and $\widetilde{\mathrm{PCC}}_{\mathrm{pd}}$ denote the min-max normalized PCC for single-crystal and powder, respectively, defined as
\[
\widetilde{\mathrm{PCC}}_{\alpha}=\frac{\mathrm{PCC}_{\alpha}-\min\limits_{\bm{\theta}\in\mathcal{E}}\mathrm{PCC}_{\alpha}}{\max\limits_{\bm{\theta}\in\mathcal{E}}\mathrm{PCC}_{\alpha}-\min\limits_{\bm{\theta}\in\mathcal{E}}\mathrm{PCC}_{\alpha}}, \qquad \alpha\in\{\mathrm{sc},\mathrm{pd}\},
\]
where $\mathcal{E}$ denotes the set of eligible samples drawn from the posterior distribution.
\fi
The corresponding Hamiltonian parameter vectors for both the reference model, $\bm{\theta}_{\mathrm{ref}}$, and the selected posterior sample, $\bm{\theta}_{\mathrm{best}}$, are listed in \ref{sec_SN:hamiltonian_param_vals}.
% [-5.94619289e-03  1.99952394e-01 -2.86565709e+00 -2.12778521e+00 1.86451375e-01  2.13745162e-01  1.36096621e+01  1.35206213e+01 -3.91540170e-01] best without binning

Figures~\ref{fig:expt_conerplot_spectrograms}(b) and (c) compare the measured and predicted powder and single-crystal spectra, respectively. For the powder INS, the left panels show the experimental $S(|\mathbf{Q}|,\omega)$ together with the spectrum generated by the ``best'' posterior sample, $S(|\mathbf{Q}|,\omega;\bm{\theta}_{\mathrm{best}})$, and that generated by the literature reference Hamiltonian, $S(|\mathbf{Q}|,\omega;\bm{\theta}_{\mathrm{ref}})$. The right panels show the corresponding spectra for the single-crystal modality along the high-symmetry path.

We emphasize here that the purpose of this comparison is not to revise the Hamiltonian parameters reported in Ref.~\citenum{scheie2023spin}, but to demonstrate how the present framework identifies alternative parameter sets that are consistent with the processed experimental spectra and characterizes the associated parameter ambiguities.

Both Hamiltonian estimates reproduce the dominant spectral features in both modalities, but differ in subtle intensity distributions and dispersive details that remain important for resolving the underlying microscopic interactions. In particular, $\bm{\theta}_{\mathrm{best}}$ yields higher PCC values for both powder and single-crystal measurements than $\bm{\theta}_{\mathrm{ref}}$, indicating that the joint multimodal posterior identifies Hamiltonians with improved simultaneous agreement with the processed experimental spectra considered here.
Interestingly, the posterior samples span a much narrower range of powder agreement than single-crystal agreement. This behavior reflects the weaker constraints imposed by powder INS: many distinct Hamiltonians produce similarly good powder spectra, whereas the directional information preserved by single-crystal INS more effectively differentiates among competing microscopic models.

Beyond returning a representative ``best'' Hamiltonian, the inference yields a full posterior distribution over Hamiltonian parameters, enabling direct quantification of the uncertainty associated with each microscopic interaction. Rather than collapsing the inference to a single point estimate, the posterior distributions and parameter correlations reveal which interaction combinations remain poorly resolved by the available measurements, consistent with the locally sloppy directions predicted by the observation geometry.

This experimental validation demonstrates that the proposed framework extends beyond simulated datasets to real multimodal INS measurements, while preserving physical interpretability through the shared Hamiltonian parameter space. The framework also provides a principled approach for determining not only which microscopic models are consistent with heterogeneous experimental observations, but also which interactions remain fundamentally unresolved by the available measurements.

\section*{Discussion}

In this work, we present an AI-enabled framework that not only accelerates Hamiltonian inference from experimentally measured spectra, but also provides a probabilistic and physically interpretable approach to Hamiltonian inference. By combining Hamiltonian-conditioned neural surrogates with Bayesian inference, local observation geometry, and Bayesian experimental design, the framework quantifies uncertainty, integrates complementary measurement modalities, and predicts the parameter-resolving capability of a given measurement configuration. 
Rather than simply producing a best-fit Hamiltonian, the framework characterizes the information content of spectroscopy measurements and the remaining ambiguities in the inferred microscopic model.

We demonstrate these capabilities using NiPS\textsubscript{3} as a representative quantum magnet with both powder and single-crystal INS measurements. The learned surrogate enables rapid, differentiable evaluation of neutron-scattering spectra throughout the Hamiltonian parameter space, making it practical to construct observation landscapes that reveal stiff and sloppy parameter directions. These landscapes provide intuitive and quantitative insight into why certain interactions are well constrained by INS whereas others remain only weakly identifiable. 
Importantly, the observation geometry itself is independent of the neural surrogate and can, in principle, be defined for any differentiable forward model. 
However, repeated evaluation of conventional forward solvers rapidly becomes computationally prohibitive in high-dimensional Hamiltonian spaces. For example, a uniform nine-dimensional grid with the same computational budget as the training dataset used in this work would provide only about three sampling points along each parameter dimension, far too coarse to accurately resolve the posterior distribution. 
The surrogate therefore serves as an enabling computational accelerator that makes repeated evaluation of the observation geometry, Bayesian inference, and adaptive experimental design practical for realistic high-dimensional Hamiltonian spaces.

A key strength of the framework is its physically grounded treatment of multimodal inference. Rather than coupling heterogeneous measurements through an abstract learned latent representation, powder and single-crystal INS are linked through their shared microscopic Hamiltonian parameter space. This enables posterior information inferred from one modality to be propagated naturally to another, supporting sequential Bayesian inference, joint parameter estimation, and adaptive experimental design. In the experimental analysis of NiPS\textsubscript{3}, this strategy yields posterior distributions that simultaneously reproduce both powder and single-crystal spectra while explicitly quantifying the remaining uncertainty.
Notably, starting from an established Hamiltonian obtained through conventional refinement, the framework identifies alternative parameter combinations consistent with the observations, providing a complementary characterization of parameter ambiguities beyond the single best-fit solution.

The present framework also has several limitations. First, the surrogate model is trained on spectra generated from an assumed Hamiltonian family and forward solver, so its predictive accuracy is ultimately limited by the fidelity of the underlying physical model, the quality of the approximation method used to calculate the spectra (here spin wave theory), and the coverage of the training dataset. Second, although neural surrogates provide highly efficient forward evaluations after training, generating sufficiently informative training data remains computationally demanding for high-dimensional Hamiltonians. Third, the inferred posterior quantifies uncertainty only within the assumed model class. Missing interactions, disorder, domain structure, instrumental effects, or other sources of model mismatch may therefore manifest as broadened posterior distributions or systematic residuals rather than being explicitly represented. Future work could address these limitations by incorporating surrogate-model uncertainty, for example through Bayesian neural networks or ensemble methods, and by using cross-modal residuals as diagnostics of model inadequacy.

Beyond Hamiltonian inference, the inferred posterior may itself encode useful physical information. Real materials often exhibit spatial inhomogeneity, strain, stacking faults, domains, or local disorder, giving rise to distributions of effective microscopic parameters rather than a single uniform Hamiltonian. Although the present work interprets the posterior primarily as epistemic uncertainty, future hierarchical Bayesian models could distinguish uncertainty arising from limited measurements from intrinsic variations within the sample. Such extensions would enable multimodal spectroscopy to probe not only the most probable Hamiltonian but also the distribution of microscopic environments contributing to the measured response.

Our framework uses AI not as an end-to-end inverse model, but as a reusable surrogate for computationally expensive physics calculations while keeping the inference grounded in microscopic Hamiltonian parameters and experimentally defined observables. 
This separation between the physical model, the learned surrogate, and the Bayesian inference engine preserves physical interpretability while enabling efficient uncertainty quantification and experimental design. 
The central requirement of this methodology, namely a computationally efficient, differentiable forward model, is shared by many spectroscopic and scattering techniques, providing a basis for applying the framework beyond INS to domains such as resonant inelastic X-ray scattering, Raman spectroscopy, X-ray absorption spectroscopy, optical spectroscopy, and transport measurements. 
By revealing how measurement modalities constrain microscopic interactions and identifying experiments that efficiently resolve remaining ambiguities, the framework establishes a general strategy for uncertainty-aware Hamiltonian inference and adaptive multimodal experimental design in quantum materials. 
Because the observation geometry is defined entirely by the forward model, the same framework naturally extends to optimizing additional experimental control variables, such as sample orientation, incident energy, magnetic field, pressure, temperature, or other external perturbations, as well as coordinating complementary measurements across different experimental modalities.
This statistical decision-making layer, combined with emerging agentic AI-driven laboratory automation~\cite{boiko2023autonomous,xiao2025crystalpilot,hellert2026agentic,chen2026agentic,qiu2026aims}, could enable closed-loop autonomous scientific experiments that iteratively identify the most informative measurements, acquire new data, and refine microscopic models.
By unifying physically interpretable AI surrogates, Bayesian inference, observation geometry, and adaptive experimental design, this work provides a foundation for autonomous scientific discovery in quantum materials.

\section*{Methods}
\label{sec:methods}
\setcounter{subsection}{0}

\subsection*{Bayesian inference and experimental design}
\label{methods:boed}

Unless otherwise specified, we adopt independent uniform priors over the nine Hamiltonian parameters,
\begin{equation}
\label{eq:uniform_prior}
P^{(0)}(\bm{\theta})
=
\prod_{m=1}^{9}
\mathcal{U}\!\left(\theta_m;\theta_{m,\mathrm{lb}},\theta_{m,\mathrm{ub}}\right),
\end{equation}
where the lower and upper bounds are
\begin{equation}
\begin{aligned}
\bm{\theta}_{\mathrm{lb}}=
(-0.020,\,0.15,\,-3.2,\,-2.2,\,0.15&,\\
0.15,\,13.5,\,13.5,\,-0.76&) \text{ and }\\
\bm{\theta}_{\mathrm{ub}}=
(-0.005,\,0.35,\,-2.5,\,-1.5,\,0.26&,\\
0.26,\,15.0,\,15.0,\,-0.25&),
\end{aligned}
\end{equation}
in the parameter order
\begin{equation}
\bm{\theta}
=
(A_x, A_z, J_{\mathrm{1a}}, J_{\mathrm{1b}}, J_{\mathrm{2a}}, J_{\mathrm{2b}}, J_{\mathrm{3a}}, J_{\mathrm{3b}}, J_4).
\end{equation}
The prior ranges were chosen to encompass physically plausible Hamiltonian values around the reference parameter set while allowing reasonable variation along each parameter direction.

The posterior distribution was approximated using a sequential Monte Carlo particle approximation,
\begin{equation}
P^{(t)}(\bm{\theta})
\approx
\sum_{n=1}^{N} w_n^{(t)} \delta(\bm{\theta}-\bm{\theta}_n^{(t)}),
\end{equation}
where $N=10{,}000$ particles were initialized by independent sampling from the prior in Eq.~\eqref{eq:uniform_prior}, with initial weights $w_n^{(0)}=1/N$. At design iteration $t$, the selected experimental setting produces a set of measured coordinates and intensities $\mathcal{D}_t=\{(\mathbf{x}_j,y_j)\}_{j=1}^{M}$. Here $\mathbf{x}_j=({H_j,K_j,L_j,\omega_j})$ for single-crystal measurements.

For each particle, the surrogate forward model produced an intensity prediction $S_{\mathrm{raw~pred}}(\mathbf{Q}_{j},\omega_{j};\bm{\theta}_{n}^{(t)})$, which was scaled and clipped for the following particle-score calculation via
\begin{equation}
    S_{\mathrm{pred}}(\mathbf{Q}_{j},\omega_{j};\bm{\theta}_{n}^{(t)})=\alpha_{t}\max\{S_{\mathrm{raw~pred}}(\mathbf{Q}_{j},\omega_{j};\bm{\theta}_{n}^{(t)}),0\}
\end{equation}
The global scale factor $\alpha_{t}$ was adaptively estimated from the ratio between the mean measured intensity and the mean particle-averaged predicted intensity, with exponential smoothing in log space. 
Particle weights were subsequently updated using a Poisson-derived robust score $\ell_{n}^{(t)}$,
\begin{equation}
w_n^{(t+1)}
=
\frac{
w_n^{(t)}\exp(\ell_n^{(t)})
}{
\sum_{m=1}^{N}w_m^{(t)}\exp(\ell_m^{(t)})
}.
\end{equation}
Because $\ell_{n}^{(t)}$ is a robust effective score rather than a conventional log-likelihood, the resulting particle distribution is interpreted as a generalized posterior.
More details about the adaptive intensity scale factor $\alpha_t$ are provided in \ref{sec_SN:global_scale_factor}, while the Poisson-derived robust score $\ell_n^{(t)}$ is described in \ref{sec_SN:robost_lkhd}.

To monitor particle degeneracy, we used the standard effective sample size
\begin{equation}
N_{\mathrm{eff}}^{(t)}=\frac{1}{\sum_{n=1}^{N} \left(w_n^{(t)}\right)^2}.
\end{equation}
When $N_{\mathrm{eff}}^{(t)}<N/2$, particles were rejuvenated using Liu--West resampling \cite{liu2001combined}. Parent particles were sampled with replacement according to the current weights, and each child particle was shrunk toward the current weighted mean and perturbed by a Gaussian kernel,
\begin{equation}
\begin{aligned}
    \bm{\theta}_n'&= a \bm{\theta}_{I_n}+(1-a)\bar{\bm{\theta}}+\bm{\epsilon}_n,
    \\
    \bm{\epsilon}_n& \sim \mathcal{N}\!\left(\bm{0},(1-a^2)\bm{\Sigma}+10^{-6}\mathbf{I}\right),
\end{aligned}
\end{equation}
where $I_n\sim\mathrm{Categorical}(\{w_m\}_{m=1}^{N})$, $\bar{\bm{\theta}}$ and $\bm{\Sigma}$ are the weighted particle mean and covariance, and $a=0.98$. After resampling, all weights were reset to $1/N$. Particles falling outside the prior bounds were replaced by new samples drawn uniformly within the allowed parameter ranges.

Because the masked powder measurements provide only weak constraints on several Hamiltonian parameters, directly propagating the entire powder particle set could result in overconfident initialization for the subsequent single-crystal inference. We therefore adopted a partial posterior initialization strategy. Specifically, particles were first resampled from the powder posterior according to their importance weights, after which the parameters $A_x$, $J_{\mathrm{1a}}$, $J_{\mathrm{1b}}$, $J_{\mathrm{2a}}$, $J_{\mathrm{2b}}$, and $J_4$ were reinitialized by independently drawing from their original uniform prior distributions, while the remaining parameters retained the powder-informed posterior values. The particle weights were subsequently reset to a uniform distribution before beginning the single-crystal inference.

Experimental settings were selected sequentially from the candidate setting space. For a candidate orientation $\psi$, we sampled coordinates $\mathbf{x}$ within the corresponding instrumental coverage region and drew parameter samples from the current particle approximation. 
The acquisition utility defined in Eq.~\eqref{eq:bed_utility} was multiplied by two scale factors to incorporate practical experimental considerations,
\begin{equation}
\tilde{U}^{(t+1)}(\psi)
=
s_{\mathrm{rep}}^{(t+1)}(\psi)\,
s_{\mathrm{dist}}^{(t+1)}(\psi)\,
U^{(t+1)}(\psi),
\end{equation}
where $s_{\mathrm{rep}}$ penalizes repeated experimental settings and $s_{\mathrm{dist}}$ penalizes large changes relative to the previously selected single-crystal orientation. The specific forms of the two scale factors are provided in \ref{sec_SN:hist_setting_penalty}. The next measurement was selected by maximizing the adjusted utility: 
\begin{equation}
\psi_{t+1}=\arg\max_\psi \tilde{U}^{(t+1)}(\psi).
\end{equation}
For the default Bayesian optimal experimental design (BOED) runs, utilities were estimated using 1000 sampled coordinates and 500 parameter samples at each design step.

\subsection*{Neural surrogate model}

We used a multimodal neural surrogate to approximate the dynamical structure
factor for both single-crystal and powder neutron-scattering geometries. The
surrogate was based on a combined feature-wise linear modulation (FiLM) sinusoidal representation network (SIREN) architecture\cite{perez2018film,sitzmann2020siren}, in which the
nine-dimensional Hamiltonian parameter vector $\bm{\theta}=[A_x,A_z,J_{\mathrm{1a}},J_{\mathrm{1b}},J_{\mathrm{2a}}, J_{\mathrm{2b}},J_{\mathrm{3a}},J_{\mathrm{3b}},J_4]$ was passed through a shared mapping network to generate layer-wise scale and shift modulations. The mapping network was a fully connected ReLU network with three hidden layers of 256 neurons, producing the FiLM parameters for the coordinate networks.

Separate coordinate networks were used for the single-crystal and powder
modalities. The single-crystal network took $(H,K,L,\omega)$ as input, while the powder network took $(|\mathbf{Q}|,\omega)$ as input. Each coordinate network consisted of four FiLM-modulated sine layers with 256 hidden units and SIREN frequency parameter $\omega_0=30$, followed by a linear output layer producing a scalar intensity. The Hamiltonian-parameter mapping network was shared between modalities, while the coordinate networks were modality-specific.

The surrogate was trained jointly on the single-crystal and powder datasets, which together comprise $20{,}000$ Hamiltonian samples (see ``Training data generation'' below). The dataset was randomly split into training, validation, and testing sets containing 90\%, 5\%, and 5\% of the samples, respectively.
We optimized a weighted mean-squared error loss with equal weights for the two modalities, using the Adam optimizer with a learning rate $10^{-4}$ and mini-batch size 4. Training was run for up to 300 epochs, and the model with the lowest validation loss (at epoch 239) was used for all BOED calculations. The detailed network architecture and training procedure are provided in \ref{sec_SN:ai_surrogate}.

\subsection*{Experiment and data preparation}

% \todo{To Drew, Matt: could you add the description of the powder sample preparation and SEQUOIA experimental details?}

Polycrystalline NiPS\textsubscript{3} was prepared by heating a mixture of high-purity elements  in a silica tube sealed under vacuum after argon purging.  The reaction occurred by slowly heating ($10\,^\circ\text{C/h}$) to $400\,^\circ\text{C}$ followed by a $24\,\text{h}$ dwell, then heating ($10\,^\circ\text{C/h}$) to $650\,^\circ\text{C}$ for $24\,\text{h}$.  The furnace power was turned off and the ampoule cooled in the furnace. This produced a mixture of crystals and polycrystalline masses, which were ground in air and passed through a $0.5\,\text{mm}$ sieve to reduce particle size.  The resulting powder was annealed for $16\,\text{h}$ at $600\,^\circ\text{C}$ in an evacuated silica tube.  The resulting material was lightly ground and passed through a $0.5\,\text{mm}$ sieve before characterization by x-ray diffraction and magnetization measurements to verify the desired phase and expected Neel temperature.  Prior to use, forming gas was used to reduce nickel powder at $450\,^\circ\text{C}$ for $12\,\text{h}$.

% \todo{Comment from Drew: I suggest you also add the mass of powder that went into the beam, if that isn't listed above..  I don't recall exactly how much i gave Daniel. It was close to 8g.}
% A $m=xxxx$~g powder sample of NiPS$_3$ was loaded under a helium atmosphere into XXX diameter aluminum sample cans.  
The powder sample of NiPS$_3$ was loaded under a helium atmosphere into aluminum sample cans.
The sample along with an identical empty can were mounted to the cold-finger of the SEQUOIA instruments closed-cycle refrigerator sample changer\cite{Stone2025}. The third position of the sample changer had no sample installed for independent measurements of the sample environment background contribution.  Inelastic neutron scattering measurements were performed at $T=5$~K for incident energy $E_i=210$ (6,3,3 C), $104.6$ (15,12,2.9 C), $62.7$ (12.3,9,1.5 C), and $29.3$ (8.9,9,1.5 C) meV with the instruments high resolution Fermi chopper spinning at 600, 540, 420, and 240 Hz respectively. Values in parentheses correspond to accumulated proton charge in Coulombs on the spallation target for measurements of the sample, the empty can, and the empty sample environment respectively. Details about extraction of powder magnon spectrum from the total experimental INS data are presented in \ref{sec_SN:extraction_expt_powder_data}.

% \todo{Daniel, Garrett, David: could you write a brief summary about how experimental powder and single-crystal data is processed?}
\textbf{Powder data.}
The powder INS data were reduced from the raw SEQUOIA event data using Mantid~\cite{Arnold2014Mantid}. For the powder spectrum used in Fig.~\ref{fig:expt_conerplot_spectrograms}(c), the nominal $E_i=100$~meV dataset was used. Individual sample runs were merged prior to reduction, with the corresponding neutron-monitor data merged separately. The incident energy and time-zero correction were determined from the monitor data, and the scattering intensity was normalized to proton charge and corrected for detector efficiency using a processed vanadium standard. The reduced detector data were transformed to powder coordinates $(|\mathbf{Q}|,\hbar\omega)$ using \texttt{ConvertToMD} and histogrammed using \texttt{BinMD} over $0\leq |\mathbf{Q}|\leq 7$~\AA$^{-1}$ and $0\leq \hbar\omega\leq 80$~meV, with bin widths of $0.035$~\AA$^{-1}$ and $0.4$~meV, respectively. Measurements of the NiPS\textsubscript{3} sample together with the aluminum sample can and heat shield were reduced independently from corresponding empty-can measurements containing the can and heat shield, using the same reduction and histogramming procedure. Rather than directly subtracting the empty-can spectrum, the sample and empty-can spectra were retained as separate inputs to the vision-transformer-based source-separation workflow described in \ref{sec_SN:extraction_expt_powder_data}. This workflow separates the magnetic scattering from sample-phonon and sample-environment contributions, after which the extracted magnon spectrum was processed by the feature-enhancement model and used for Hamiltonian inference.

\textbf{Single-crystal data.}
The experimental spectrum along the high-symmetry path shown in Fig.~\ref{fig:expt_conerplot_spectrograms}(c) was generated from the original Mantid MDEventWorkspace (MDE) files~\cite{Arnold2014Mantid} used to produce the figures in Ref.~\citenum{scheie2023spin}. The dataset comprises three merged MDE files containing complete sample-rotation measurements collected at incident energies $E_i=28$, 60, and 100~meV. For each incident energy, the data were histogrammed along the high-symmetry path $C-\Gamma$, $\Gamma-Y$, $Y-C$, $C-Z$, and $Z-\Gamma$ using the MDNorm algorithm as implemented in SHIVER~\cite{Savici2022,Savici2025SHIVER}. The histogram specifications followed Ref.~\citenum{scheie2023spin}, with integration widths of $\pm0.05$ reciprocal lattice units (r.l.u.) in the in-plane direction transverse to the path and $\pm0.25$~r.l.u. in the out-of-plane direction. An energy-bin width of $0.5$~meV was used over $0\leq\hbar\omega\leq60$~meV. The individual path segments were concatenated for each incident energy and then combined into a single spectrum, using the $E_i=28$~meV data for $0$--$20$~meV, the $E_i=60$~meV data for $20$--$32$~meV, and the $E_i=100$~meV data above $32$~meV. The effect of the finite reciprocal-space integration used to construct these cuts on the Hamiltonian inference is examined further in \ref{sec_SN:finite_q_binning}.
%All single-crystal measurements were performed under {\color{red}IPTS-37598} on the SEQUOIA instrument at the Spallation Neutron Source, Oak Ridge National Laboratory. 

\subsection*{Training data generation}

Training data were generated using linear spin wave calculations implemented in \texttt{Sunny.jl}~\cite{dahlbom2025sunny}. We randomly sampled $20{,}000$ Hamiltonians by independently varying each Hamiltonian parameter over a range between zero and twice its value reported in Ref.~\citenum{scheie2023spin}, while preserving its sign; i.e., $\theta_i \sim \mathcal{U}(0,\,2\theta_{\mathrm{ref},i})$ for $\theta_{\mathrm{ref},i}>0$ and $\theta_i \sim \mathcal{U}(2\theta_{\mathrm{ref},i},\,0)$ for $\theta_{\mathrm{ref},i}<0$. For each parameter set, the magnetic ground state was first obtained through classical energy minimization before constructing the corresponding spin wave model. The magnetic ground state search was always conducted on a single crystallographic unit cell. Parameter sets that failed to converge or yielded unstable magnon spectra were discarded and replaced with new samples.

Single-crystal training data were generated by evaluating the dynamical structure factor at $2{,}000$ randomly sampled momentum-transfer points uniformly distributed within $[-3,3]^3$ reciprocal lattice units (r.l.u.) and $100$ uniformly spaced energy transfers between $0$ and $100$ meV. Powder spectra were generated using orientational powder averaging with $300$ randomly sampled $|\mathbf{Q}|$ values and $3{,}000$ crystal orientations. Both single-crystal and powder spectra employed Gaussian energy broadening with a full width at half maximum (FWHM) of $4$ meV. Three rotational magnetic domains were averaged for the single-crystal calculations, and the Ni$^{2+}$ magnetic form factor was included throughout. The Hamiltonian parameters and scattering coordinates were linearly normalized to the interval $[-1,1]$ before training.

\section*{Data availability}

The experimental data supporting the findings of this study are available from the corresponding author upon reasonable request. The simulation datasets can be regenerated using the code provided and will be made publicly available upon acceptance of the manuscript.

\section*{Code availability}

The source code required to reproduce the results presented in this work will be made publicly available through a GitHub repository with a permanent Zenodo archive upon acceptance of the manuscript.

\begin{acknowledgments}
This work is primarily supported by the U.S. Department of Energy, Office of Science, Basic Energy Sciences (BES) under the Genesis Mission BES AI Pathfinder Program, MAIQMag: Multimodal AI for 2D Quantum Magnets.
% \todo{To everyone: please add your acknowledgement.}
Sample synthesis and verification (A.F.M.) were supported by the U. S. Department of Energy, Office of Science, Basic Energy Sciences, Materials Sciences and Engineering Division.
This research used computational resources of the National Energy Research Scientific Computing Center (NERSC), a US Department of Energy Office of Science User Facility located at Lawrence Berkeley National Laboratory operated under contract DE-AC02-05CH11231, using NERSC award ERCAP0038195.
The authors acknowledge the Texas Advanced Computing Center (TACC) at The University of Texas at Austin for providing computational resources under project code DMR25014 that have contributed to the research results reported within this paper. We thank Allen Scheie for providing the single-crystal INS data originally published in Ref.~\citenum{scheie2023spin}.
A portion of this research used resources at the Spallation Neutron Source, a DOE Office of Science User Facility operated by the Oak Ridge National Laboratory. The beam time was allocated to SEQUOIA on proposal number IPTS-37598.
During the preparation of this work, the authors used the large language model ChatGPT by OpenAI in order to refine the language and enhance the readability of this paper. After using this tool, the authors reviewed and edited the content as needed and take full responsibility for all content in this publication.
\end{acknowledgments}

\bibliography{references}

\clearpage

\onecolumngrid
\appendix
\renewcommand{\thesubsection}{Supplementary Note \arabic{subsection}}
\setcounter{subsection}{0}

% Figures
\renewcommand{\thefigure}{S\arabic{figure}}
\setcounter{figure}{0}

% Tables
\renewcommand{\thetable}{S\arabic{table}}
\setcounter{table}{0}

% Equations
\renewcommand{\theequation}{S\arabic{equation}}
\setcounter{equation}{0}

\setcounter{page}{1}

\section*{}

\subsection{Global scale factor}
\label{sec_SN:global_scale_factor}

For each particle, the surrogate forward model produced an intensity prediction $S_{\mathrm{pred}}(\mathbf{Q}_{j},\omega_{j};\bm{\theta}_{n}^{(t)})$. Since the surrogate intensity scale need not match the measured count scale, we introduced an adaptive global scale factor, where $M$ is the number of measured data points,
\begin{equation}
\widehat{\alpha}_t = \frac{M^{-1}\sum_{j=1}^{M}\max(S_{\mathrm{targ}}(\mathbf{Q}_{j},\omega_{j}),0)}{M^{-1}\sum_{j=1}^{M}N^{-1}\sum_{n=1}^{N}\max\{S_{\mathrm{raw~pred}}(\mathbf{Q}_{j},\omega_{j};\bm{\theta}_{n}^{(t)}),0\}
+\varepsilon},\qquad \varepsilon=10^{-12}.
\end{equation}
The scale used in the particle score, $\alpha_t$, was obtained by smoothing $\widehat{\alpha}_t$ with an exponential moving average in log space,
\begin{equation}
\log \alpha_t = \rho \log \alpha_{t-1} + (1-\rho)\log \widehat{\alpha}_t,
\end{equation}
with $\rho=0.95$, except for the first update where $\alpha_t=\widehat{\alpha}_t$. The scaled surrogate prediction becomes
\begin{equation}
S_{\mathrm{pred}}(\mathbf{Q}_{j},\omega_{j};\bm{\theta}_{n}^{(t)}) = \alpha_t \max\left[S_{\mathrm{raw~pred}}(\mathbf{Q}_{j},\omega_{j};\bm{\theta}_{n}^{(t)}),0\right].
\end{equation} 

\clearpage
\subsection{Poisson-derived robust score}
\label{sec_SN:robost_lkhd}

We use a Poisson-derived robust particle score to define a generalized Bayesian update rather than directly evaluating the raw Poisson likelihood, which was found to be numerically too sharp for the sparse, high-intensity spectra considered here.
We first computed
\begin{equation}
r_{nj}^{(t)}
=
\log \operatorname{Pois}\!\left(S_{\mathrm{targ}}(\mathbf{Q}_{j},\omega_{j});S_{\mathrm{pred}}(\mathbf{Q}_{j},\omega_{j};\bm{\theta}_{n}^{(t)})\right).
\end{equation}
For numerical stabilization, these scores were normalized over the measured data points for each particle,
\begin{equation}
p_{nj}^{(t)}
=
\frac{\exp(r_{nj}^{(t)})}
{\sum_{k=1}^{M}\exp(r_{nk}^{(t)})},
\end{equation}
and then rescaled using min--max normalization across particles at each measured point,
\begin{equation}
\tilde{p}_{nj}^{(t)}
=
\frac{
p_{nj}^{(t)}-\min_m p_{mj}^{(t)}
}{
\max_m p_{mj}^{(t)}-\min_m p_{mj}^{(t)}+\varepsilon
}
+\varepsilon,\qquad \varepsilon=10^{-12}.
\end{equation}
The score used for the particle update was
\begin{equation}
\ell_n^{(t)}
=
\frac{1}{M}\sum_{j=1}^{M}\log \tilde{p}_{nj}^{(t)}.
\end{equation}
% Weights were updated in log space according to Bayes' rule and then normalized,
The resulting score is not a conventional log-likelihood because of the normalization and rescaling operations above. We therefore use it as an effective log-score in a generalized Bayesian particle update,
\begin{equation}
w_n^{(t+1)}
=
\frac{
w_n^{(t)}\exp(\ell_n^{(t)})
}{
\sum_{m=1}^{N}w_m^{(t)}\exp(\ell_m^{(t)})
}.
\end{equation}
Accordingly, the resulting particle distribution should be interpreted as an effective (generalized) posterior induced by this robust score rather than as the posterior associated with the raw Poisson likelihood.

\clearpage
\subsection{History-dependent setting penalties}
\label{sec_SN:hist_setting_penalty}

The BOED acquisition utility was multiplied by two history-dependent penalty factors,
\begin{equation}
\tilde{U}^{(t+1)}(\psi)
=
s_{\mathrm{rep}}^{(t+1)}(\psi)
s_{\mathrm{dist}}^{(t+1)}(\psi)
U^{(t+1)}(\psi),
\end{equation}
where $s_{\mathrm{rep}}$ discourages repeated measurements at the same setting
and $s_{\mathrm{dist}}$ discourages abrupt jumps in the single-crystal rotation
angle. These factors were introduced as pragmatic scheduling regularizers: they
do not modify the posterior update, but only bias the selection of the next
experimental setting among candidates with similar acquisition utilities.

Let $\mathcal H^{(t)}=\{\psi^{(1)},\ldots,\psi^{(t)}\}$ denote the history of selected
angles before step $t+1$. The repetition penalty was defined as
\begin{equation}
s_{\mathrm{rep}}^{(t+1)}(\psi)
=
\exp\left[-\gamma_{\mathrm{rep}} C^{(t)}(\psi)\right],
\end{equation}
where
\begin{equation}
C^{(t)}(\psi)
=
\sum_{k=1}^{t}
\delta\!\left(\psi-\psi^{(k)}\right)
% \mathbf{1}\!\left(|\psi-\psi_k|<\epsilon_{\psi}\right)
\end{equation}
counts the number of previous selections of angle $\psi$, within a
small numerical tolerance $\epsilon_{\psi}$. In the calculations reported here,
$\gamma_{\mathrm{rep}}=0.5$. Thus, each repeated use of the same angle reduces its acquisition score by a factor of $\exp(-0.5)$.

For single-crystal BOED, we also applied a distance penalty relative to the most
recently selected angle $\psi^{(t)}$,
\begin{equation}
s_{\mathrm{dist}}^{(t+1)}(\psi)
=
\exp\left[
-\gamma_{\mathrm{dist}}
\left(
\frac{d_{\mathrm{per}}(\psi,\psi^{(t)})}{\ell_{\psi}}
\right)^2
\right].
\end{equation}
Here $d_{\mathrm{per}}$ is the periodic angular distance,
\begin{equation}
d_{\mathrm{per}}(\psi,\psi^{(t)})
=
\min\left(
|\psi-\psi^{(t)}|,
360^\circ-|\psi-\psi^{(t)}|
\right),
\end{equation}
to account for the $360^\circ$ periodicity of the sample rotation. We used
$\gamma_{\mathrm{dist}}=0.1$ and $\ell_{\psi}=15^\circ$. At the first design
step, before any setting has been selected, $s_{\mathrm{dist}}=1$ for all
candidate angles.

The final multiplicative penalty factor was clipped from below,
\begin{equation}
s_{\mathrm{rep}}^{(t+1)}(\psi)
s_{\mathrm{dist}}^{(t+1)}(\psi)
\leftarrow
\max\left[
s_{\mathrm{rep}}^{(t+1)}(\psi)
s_{\mathrm{dist}}^{(t+1)}(\psi),
10^{-6}
\right],
\end{equation}
to avoid assigning exactly zero acquisition score to any candidate setting. The
next setting was then chosen by maximizing the penalized acquisition utility,
\begin{equation}
\psi^{t+1}
=
\arg\max_{\psi}
\tilde{U}^{(t+1)}(\psi).
\end{equation}

\clearpage
\subsection{AI surrogate architecture and training}
\label{sec_SN:ai_surrogate}

The multimodal surrogate consists of a shared Hamiltonian-parameter mapping network and separate coordinate networks for the single-crystal and powder modalities. The mapping network takes the nine-dimensional Hamiltonian parameter vector $\bm{\theta}$ as input to produce the FiLM modulation parameters used by the modality-specific coordinate networks. The single-crystal branch takes four-dimensional coordinates $(H,K,L,\omega)$ as input, whereas the powder branch takes two-dimensional coordinates $(|\mathbf{Q}|,\omega)$ as input. Both branches use four modulated sine layers with 256 hidden units, followed by a linear scalar-output layer. The complete model architecture, as reported
by the implementation, is shown below.

{\footnotesize
\begin{verbatim}
MultiModalModulatedSiren(
  (coordinate_networks): ModuleDict(
    (single_crystal): Sequential(
      (0): ModulatedSineLayer(
        (linear): Linear(in_features=4, out_features=256, bias=True)
      )
      (1): ModulatedSineLayer(
        (linear): Linear(in_features=256, out_features=256, bias=True)
      )
      (2): ModulatedSineLayer(
        (linear): Linear(in_features=256, out_features=256, bias=True)
      )
      (3): ModulatedSineLayer(
        (linear): Linear(in_features=256, out_features=256, bias=True)
      )
      (4): Linear(in_features=256, out_features=1, bias=True)
    )
    (powder): Sequential(
      (0): ModulatedSineLayer(
        (linear): Linear(in_features=2, out_features=256, bias=True)
      )
      (1): ModulatedSineLayer(
        (linear): Linear(in_features=256, out_features=256, bias=True)
      )
      (2): ModulatedSineLayer(
        (linear): Linear(in_features=256, out_features=256, bias=True)
      )
      (3): ModulatedSineLayer(
        (linear): Linear(in_features=256, out_features=256, bias=True)
      )
      (4): Linear(in_features=256, out_features=1, bias=True)
    )
  )
  (mapping_net): MappingNetwork(
    (net): Sequential(
      (0): Linear(in_features=9, out_features=256, bias=True)
      (1): ReLU(inplace=True)
      (2): Linear(in_features=256, out_features=256, bias=True)
      (3): ReLU(inplace=True)
      (4): Linear(in_features=256, out_features=256, bias=True)
      (5): ReLU(inplace=True)
      (6): Linear(in_features=256, out_features=2048, bias=True)
    )
  )
)
\end{verbatim}
}

\begin{figure}
    \centering
    \includegraphics[width=0.4\linewidth]{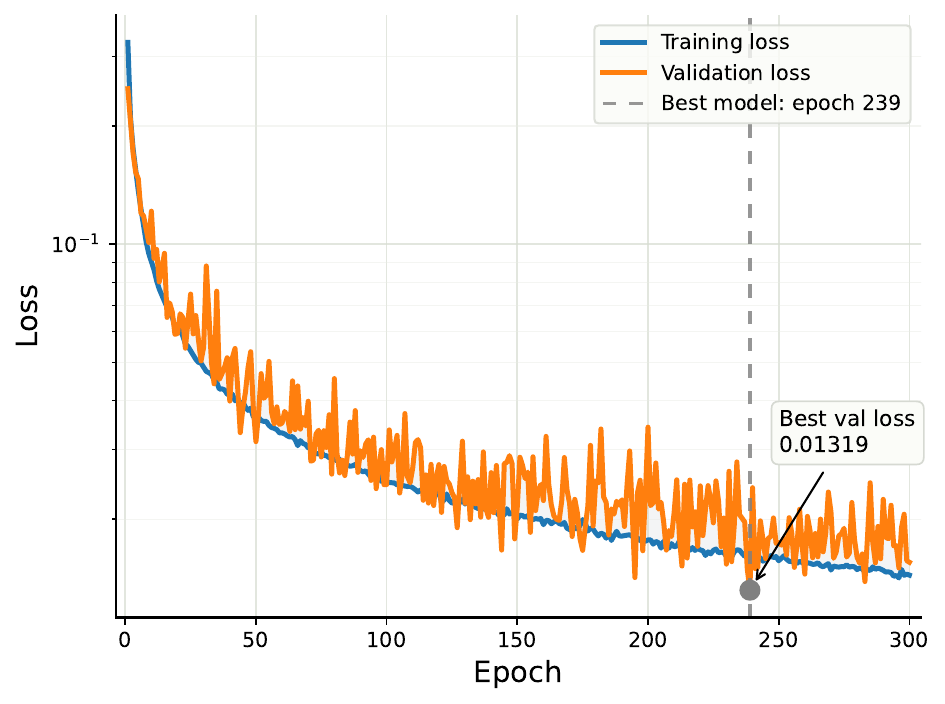}
    \caption{
    \textbf{Training and validation loss histories of the multimodal neural surrogate.}
    Training and validation losses are shown as functions of training epoch on a logarithmic scale. The model was trained for 300 epochs. The dashed vertical line marks the selected checkpoint at epoch 239, which achieved the minimum validation loss of $0.01319$. This checkpoint was used for the subsequent Hamiltonian inference and Bayesian experimental-design calculations.
    }
    \label{fig:training_validation_loss_history}
\end{figure}

The surrogate was trained for 300 epochs. Figure~\ref{fig:training_validation_loss_history} shows the training and validation loss histories on a logarithmic scale. Both losses decrease rapidly during the initial stage of training and continue to improve more gradually at later epochs. The validation loss exhibits larger epoch-to-epoch fluctuations than the training loss, as expected, but does not show sustained divergence from the training curve. The checkpoint with the lowest validation loss was obtained at epoch 239 and was used in the subsequent inference and experimental-design calculations.

\clearpage
\subsection{Hamiltonian parameter values}
\label{sec_SN:hamiltonian_param_vals}

Table~\ref{tab:hamiltonian_parameters} lists the Hamiltonian parameters of the literature reference model, $\bm{\theta}_{\rm{ref}}$, and the selected posterior sample, $\bm{\theta}_{\rm{best}}$, reported in the main text.

\begin{table}[htbp]
\caption{
Reference and inferred Hamiltonian parameters (meV).
The reference Hamiltonian is taken from Ref.~\citenum{scheie2023spin},
while $\bm{\theta}_{\rm{best}}$ denotes the posterior sample selected
according to Eq.~\eqref{eq:theta_best}.
}
\label{tab:hamiltonian_parameters}
\centering
\begin{tabular}{lccccccccc}
\hline
 & $A_x$ & $A_z$ &
 $J_{\mathrm{1a}}$ & $J_{\mathrm{1b}}$ &
 $J_{\mathrm{2a}}$ & $J_{\mathrm{2b}}$ &
 $J_{\mathrm{3a}}$ & $J_{\mathrm{3b}}$ &
 $J_{\mathrm{4}}$ \\
\hline
$\bm{\theta}_{\rm{ref}}$
&
$-0.005$
&
$0.21$
&
$-2.70$
&
$-2.00$
&
$0.20$
&
$0.20$
&
$13.90$
&
$13.90$
&
$-0.38$
\\

$\bm{\theta}_{\mathrm{best}}$
&
$-0.00595$
&
$0.20$
&
$-2.87$
&
$-2.13$
&
$0.19$
&
$0.21$
&
$13.61$
&
$13.52$
&
$-0.39$
\\
\hline
\end{tabular}
\end{table}

\clearpage
\subsection{Convergence rate comparison: sequential, random, and BOED}
\label{sec_SN:convergence_comparison}

We compare the convergence behavior of three data-acquisition strategies: sequential acquisition, random acquisition, and BOED. For each strategy, we track the mean absolute error (MAE) of the inferred Hamiltonian parameters as a function of the acquisition step. The benchmark includes ten trials in total: nine Hamiltonian parameter sets randomly sampled from the prior distribution $P_{0}(\bm{\theta})$, together with the reference parameter set reported in Ref.~\citenum{scheie2023spin}.

\begin{figure}[htbp]
    \centering
    \includegraphics[width=0.5\linewidth]{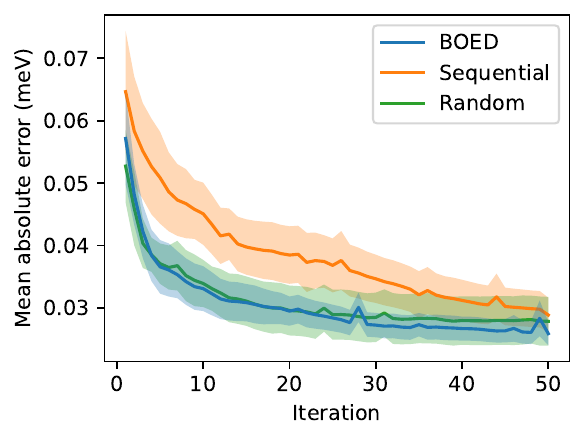}
    \caption{
    Comparison of parameter-estimation error histories for sequential, random, and BOED acquisition strategies. Solid curves show the mean absolute error (MAE) averaged over ten benchmark trials, consisting of nine randomly sampled Hamiltonian parameter sets from the prior $P_{0}(\bm{\theta})$ and one reference parameter set reported in Ref.~\citenum{scheie2023spin}. Shaded regions indicate $\pm 1$ standard deviation across trials. BOED converges substantially faster than the sequential strategy and achieves comparable or lower final error with reduced variability relative to random acquisition.
    }
    \label{fig:mae_comparison}
\end{figure}

As shown in Fig.~\ref{fig:mae_comparison}, BOED substantially accelerates convergence compared to the sequential strategy, reaching a low-error regime within the first few measurements. In contrast, the sequential strategy decreases the error more gradually, indicating that a fixed acquisition order can spend measurements on configurations that provide limited additional information.

Random acquisition achieves a mean performance that is relatively close to BOED in this benchmark, but with noticeably greater variability across trials. This behavior is expected because random acquisition may occasionally sample informative configurations, but does not consistently target the measurements that are most informative under the current posterior uncertainty. We also note that fully random acquisition is rarely used as a practical experimental protocol, where measurement choices are typically constrained by time, prior knowledge, and instrument operation. BOED therefore provides a more systematic and reproducible strategy for rapidly reducing Hamiltonian uncertainty under a finite measurement budget.

\clearpage
\subsection{High-dimensional eigen-structure}
\label{sec_SN:hidim_eigen}

\begin{figure}[htbp]
    \centering
    \includegraphics[width=0.7\linewidth]{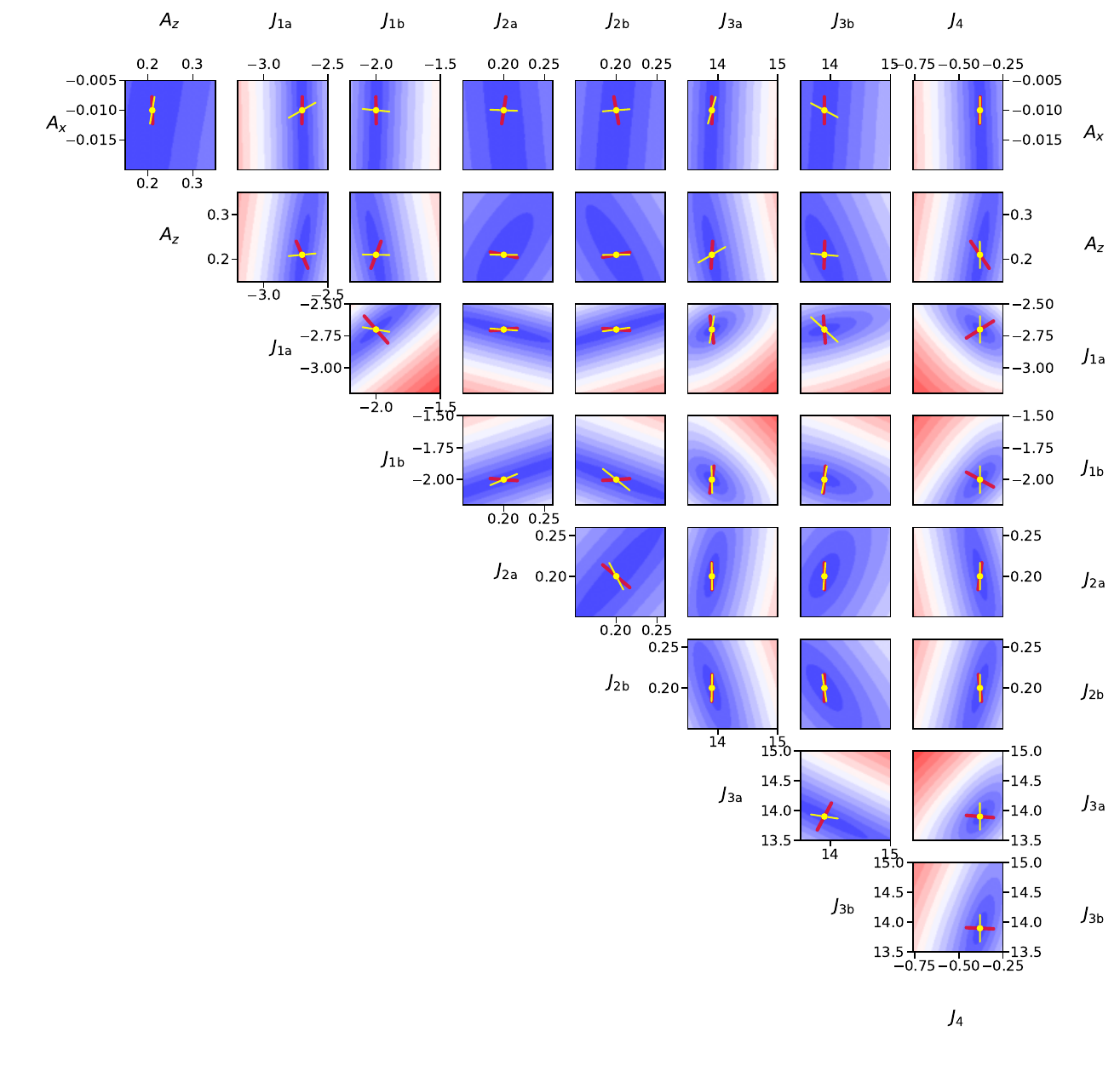}
    \caption{
    Pairwise spectral discrepancy landscapes for the single-crystal modality with projections of the global nine-dimensional metric tensor eigen-directions. Red and yellow line segments denote the projected stiffest (largest eigenvalue) and sloppiest (smallest eigenvalue) eigenvectors, respectively. Since each panel represents a two-dimensional projection of the full nine-dimensional parameter space, the projected eigen-directions are not generally expected to align with the principal axes of the corresponding two-dimensional discrepancy landscapes.
    }
    \label{fig:sc_landscape_9d}
\end{figure}

\begin{figure}
    \centering
    \includegraphics[width=0.7\linewidth]{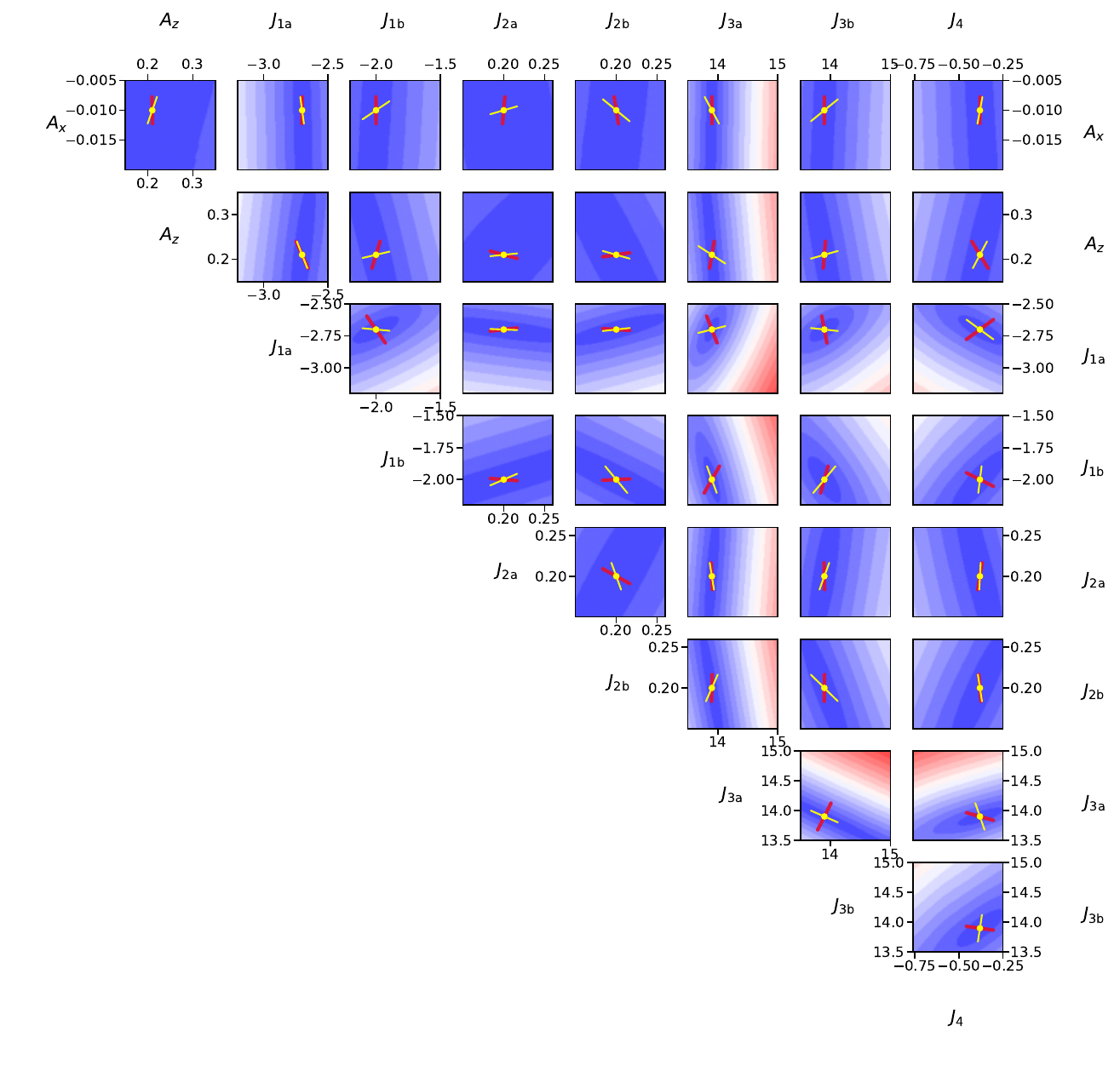}
    \caption{
    Pairwise spectral discrepancy landscapes for the powder modality with projections of the global nine-dimensional metric tensor eigen-directions. Red and yellow line segments denote the projected stiffest (largest eigenvalue) and sloppiest (smallest eigenvalue) eigenvectors, respectively. Similar to the single-crystal case, the projected global eigen-directions are not generally expected to align with the principal axes of the two-dimensional discrepancy landscapes because they represent projections of directions defined in the full nine-dimensional parameter space.
    }
    \label{fig:powder_landscape_9d}
\end{figure}

To further illustrate the geometric interpretation of the metric tensor, Fig.~\ref{fig:sc_landscape_9d} shows pairwise spectral discrepancy landscapes for the single-crystal modality, overlaid with the projections of the global stiffest (red, largest eigenvalue) and sloppiest (yellow, smallest eigenvalue) eigenvectors obtained from the full nine-dimensional metric tensor. Similar results for the powder modality are presented in Fig.~\ref{fig:powder_landscape_9d}.

Several observations are worth noting. First, the projected eigenvectors should be interpreted with caution because each panel represents only a two-dimensional projection of a nine-dimensional eigen-direction. Furthermore, the parameters span substantially different numerical ranges, producing unequal axis scales that can visually distort the projected directions.
Nevertheless, the projected stiffest direction remains approximately perpendicular to the elongated discrepancy valleys for the most sensitive parameter pairs, such as the $J_{\mathrm{1a}}$--$J_{\mathrm{1b}}$ and $J_{\mathrm{3a}}$--$J_{\mathrm{3b}}$ subspaces, consistent with these parameter pairs contributing strongly to the largest-curvature directions of the global metric.

\begin{figure}
    \centering
    \includegraphics[width=0.5\linewidth]{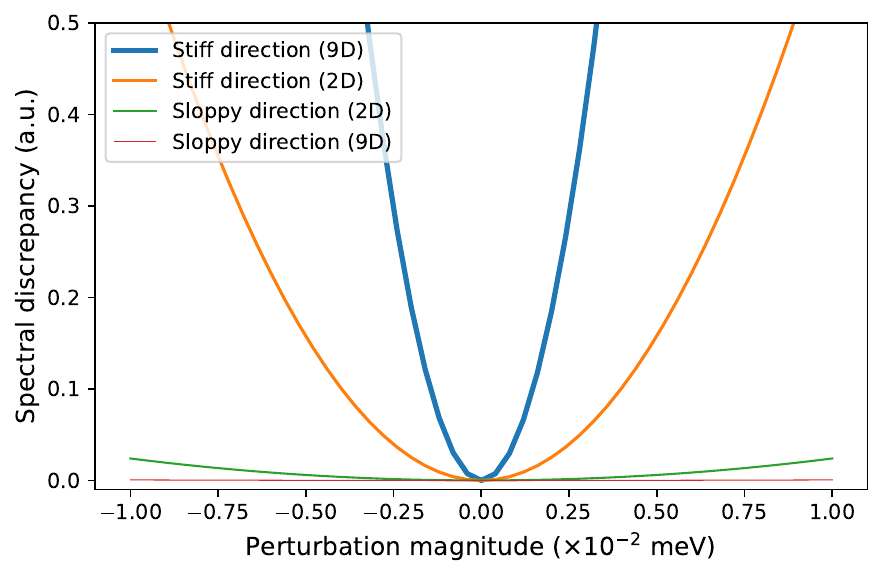}
    \caption{
    Comparison of spectral discrepancy along the global nine-dimensional and local two-dimensional principal directions for the single-crystal modality. The blue and red curves correspond to perturbations along the stiffest and sloppiest eigenvectors of the full nine-dimensional metric tensor, respectively, while the orange and green curves correspond to the principal directions obtained from the $J_{\mathrm{1a}}$--$J_{\mathrm{1b}}$ two-dimensional subspace. The global stiff and sloppy directions exhibit the fastest and slowest discrepancy growth, respectively, demonstrating that the full metric tensor eigen-directions bound the local sensitivity observed within lower-dimensional parameter subspaces.
    }
    \label{fig:sc_stiff_sloppy_comparison}
\end{figure}

\begin{figure}
    \centering
    \includegraphics[width=0.5\linewidth]{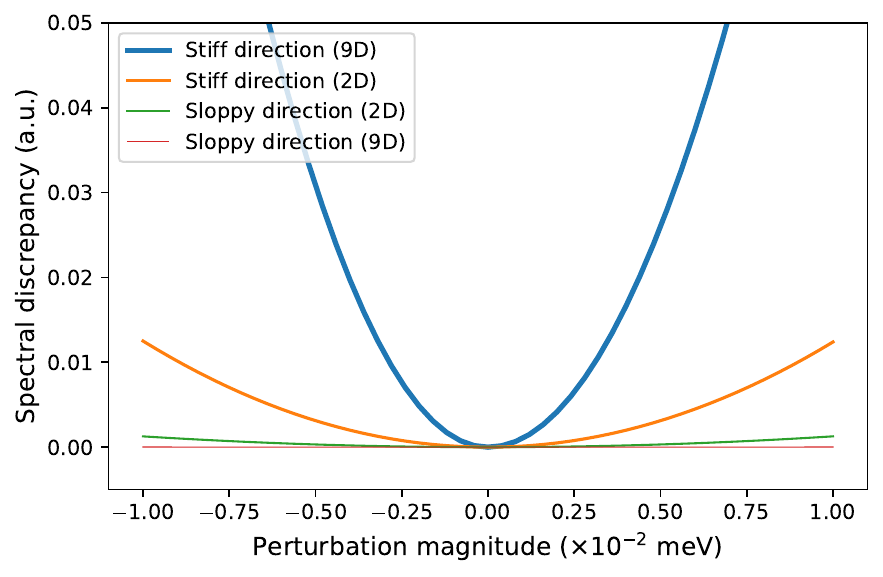}
    \caption{
    Comparison of spectral discrepancy along the global nine-dimensional and local two-dimensional principal directions for the powder modality. The global stiff and sloppy eigen-directions exhibit the fastest and slowest discrepancy growth, respectively, while the corresponding $J_{\mathrm{1a}}$--$J_{\mathrm{1b}}$ two-dimensional principal directions show intermediate behavior, consistent with the observations for the single-crystal modality.
    }
    \label{fig:powder_stiff_sloppy_comparison}
\end{figure}

To directly validate the eigen-decomposition, Fig.~\ref{fig:sc_stiff_sloppy_comparison} compares the spectral discrepancy along the global nine-dimensional stiff and sloppy eigenvectors with the corresponding principal directions obtained from the $J_{\mathrm{1a}}$--$J_{\mathrm{1b}}$ two-dimensional subspace. The discrepancy increases most rapidly along the global stiff direction and most slowly along the global sloppy direction, while the two-dimensional principal directions exhibit intermediate behavior. This demonstrates that the full nine-dimensional eigen-directions provide upper and lower bounds on the local sensitivity observed within individual parameter subspaces. A similar trend is observed for the powder modality (Fig.~\ref{fig:powder_stiff_sloppy_comparison}), which is expected because powder INS provides an orientationally averaged measurement of the single-crystal response.

\clearpage
\subsection{Pixel-wise decomposition of spectral correlation}
\label{sec_SN:pixelwise_pcc}

\begin{figure}[htbp]
    \centering
    \subfloat[]{
        \includegraphics[width=0.45\textwidth]
        {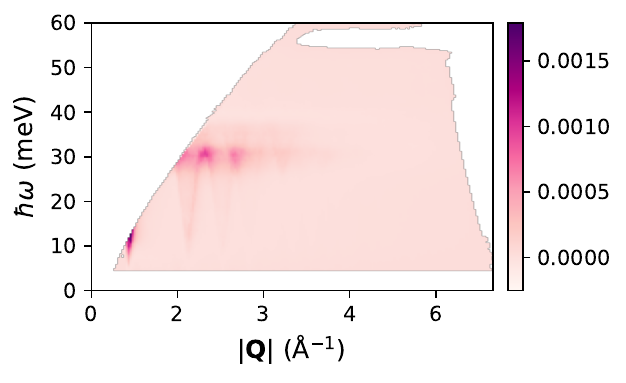}
        \label{fig:powder_pcc_contribution_best_wo_binning}
    }
    \hfill
    \subfloat[]{
        \includegraphics[width=0.45\textwidth]
        {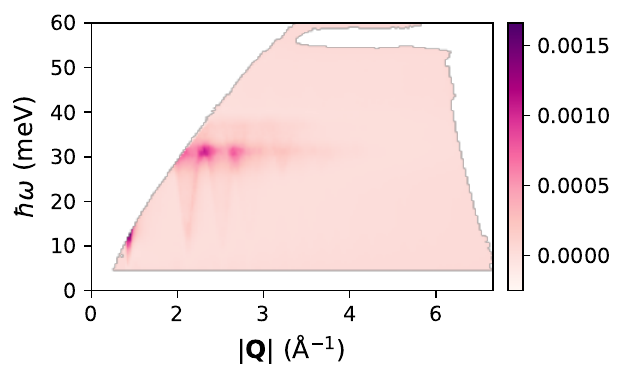}
        \label{fig:powder_pcc_contribution_ref_wo_binning}
    }
    \caption{
    Pixel-wise decomposition of the powder spectra Pearson correlation coefficient, $\mathrm{PCC}_{\mathrm{pd}}$, for
    \textbf{(a)} the selected Hamiltonian $\bm{\theta}_{\mathrm{best}}$ and
    \textbf{(b)} the literature reference Hamiltonian $\bm{\theta}_{\mathrm{ref}}$.
    }
    \label{fig:powder_pcc_contribution_wo_binning}
\end{figure}

\begin{figure}[htbp]
    \centering
    \subfloat[]{
        \includegraphics[width=0.45\textwidth]
        {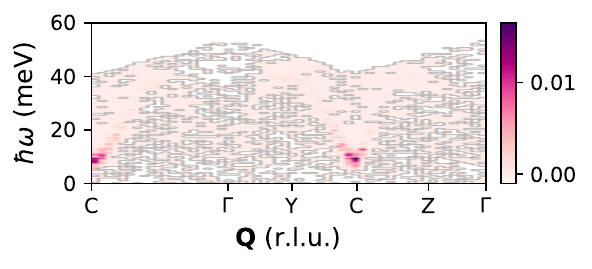}
        \label{fig:sc_pcc_contribution_best_wo_binning}
    }
    \hfill
    \subfloat[]{
        \includegraphics[width=0.45\textwidth]
        {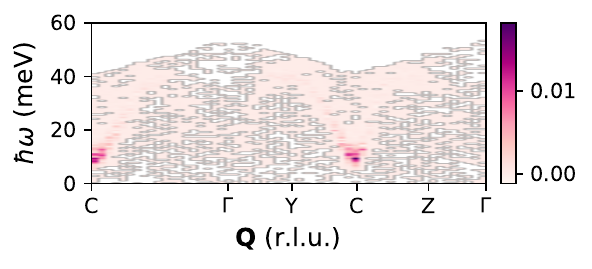}
        \label{fig:sc_pcc_contribution_ref_wo_binning}
    }
    \caption{
    Pixel-wise decomposition of the single-crystal spectra Pearson correlation coefficient, $\mathrm{PCC}_{\mathrm{sc}}$, for
    \textbf{(a)} the selected Hamiltonian $\bm{\theta}_{\mathrm{best}}$ and
    \textbf{(b)} the literature reference Hamiltonian $\bm{\theta}_{\mathrm{ref}}$.
    }
    \label{fig:sc_pcc_contribution_wo_binning}
\end{figure}

To identify the spectral regions responsible for the observed agreement, we decompose the Pearson correlation coefficient into additive pixel-wise contributions. For either modality $\alpha\in\{\mathrm{pd},\mathrm{sc}\}$, let $x_i^{(\alpha)}$ and $y_i^{(\alpha)}$ denote the predicted and experimental intensities, respectively, at valid pixel $i$. The corresponding Pearson correlation coefficient can be written as
\begin{equation}
\mathrm{PCC}_{\alpha} = \frac{ \sum_i \left(x_i^{(\alpha)}-\bar{x}^{(\alpha)}\right)\left(y_i^{(\alpha)}-\bar{y}^{(\alpha)}\right)}{\sqrt{\sum_i\left(x_i^{(\alpha)}-\bar{x}^{(\alpha)}\right)^2}\sqrt{\sum_i\left(y_i^{(\alpha)}-\bar{y}^{(\alpha)}\right)^2}}=\sum_i c_i^{(\alpha)},
\end{equation}
where
\begin{equation}
c_i^{(\alpha)}=\frac{\left(x_i^{(\alpha)}-\bar{x}^{(\alpha)}\right)\left(y_i^{(\alpha)}-\bar{y}^{(\alpha)}\right)}{\sqrt{\sum_j\left(x_j^{(\alpha)}-\bar{x}^{(\alpha)}\right)^2}\sqrt{\sum_j\left(y_j^{(\alpha)}-\bar{y}^{(\alpha)}\right)^2}}
\end{equation}
is the contribution of the $i$th pixel to the global correlation $\mathrm{PCC}_{\alpha}$.
Figures~\ref{fig:powder_pcc_contribution_wo_binning} and~\ref{fig:sc_pcc_contribution_wo_binning} show the pixel-wise contribution maps for the powder and single-crystal spectra, respectively, comparing the selected Hamiltonian $\bm{\theta}_{\mathrm{best}}$ and the reference Hamiltonian $\bm{\theta}_{\mathrm{ref}}$ with the experimental data. 
Larger positive contributions identify spectral regions that support the global correlation, whereas negative contributions identify regions that reduce it. These maps reveal which spectral regions account for the differences in global agreement reported in Fig.~\ref{fig:expt_conerplot_spectrograms}.

\clearpage
\subsection{Effect of finite momentum-space binning}
\label{sec_SN:finite_q_binning}

The single-crystal experimental spectra along the high-symmetry path were obtained by integrating the measured intensity over finite regions of reciprocal space. Directly incorporating this binning procedure during random-coordinate training data generation would substantially increase the computational cost. 
We therefore use the same sharp-coordinate surrogate and subsequently average its predictions over the corresponding experimental momentum-bin volumes. This separation also preserves the generality of the surrogate, allowing different momentum-space resolutions and integration widths to be applied at inference time without retraining. We do not apply additional energy binning because the surrogate is trained on simulated spectra with an energy broadening of $\mathrm{FWHM}=4\,\mathrm{meV}$.

\begin{figure}[htbp]
    \centering
    \includegraphics[width=0.85\linewidth]{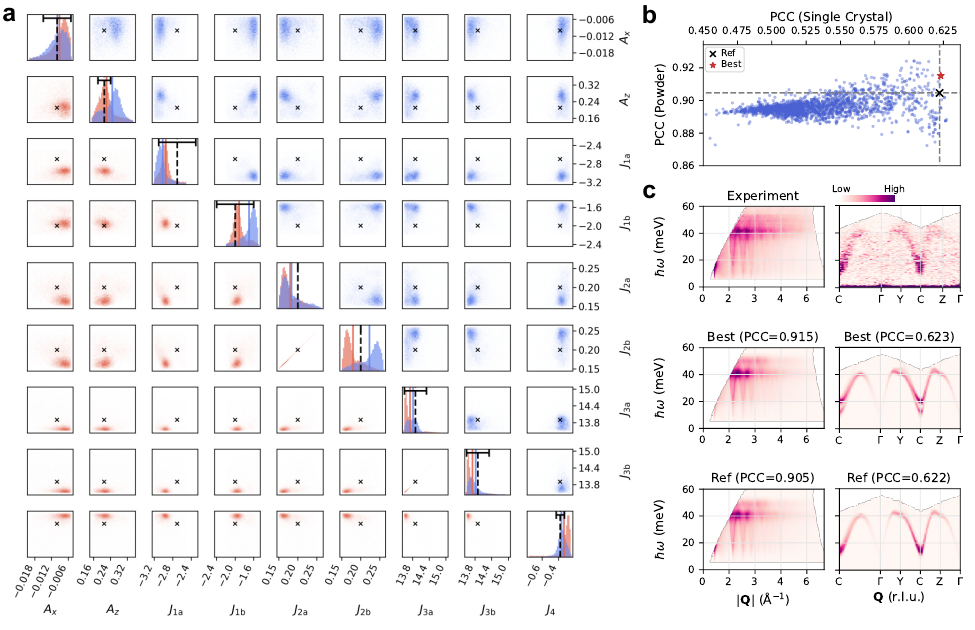}
    \caption{
    Finite $\mathbf{Q}$-space binning comparison corresponding to Fig.~\ref{fig:expt_conerplot_spectrograms}. The single-crystal surrogate spectra are averaged over the experimental momentum-bin volumes during Bayesian inference to obtain the reported posterior distribution. All panel definitions, symbols, and plotting conventions follow Fig.~\ref{fig:expt_conerplot_spectrograms}; see its caption for details.
    }
    \label{fig:expt_conerplot_spectrograms_w_binning}
\end{figure}

\begin{figure}[htbp]
    \centering
    \subfloat[]{
        \includegraphics[width=0.45\textwidth]
        {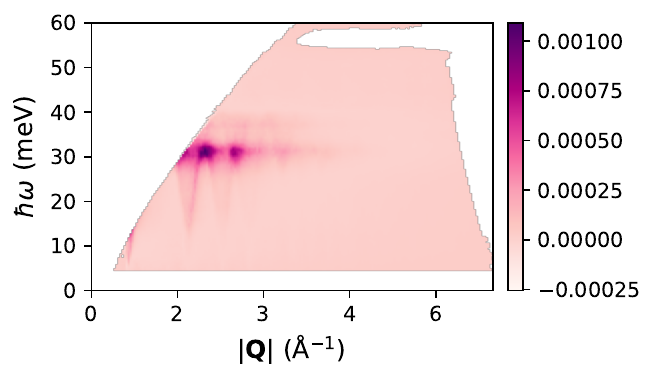}
        \label{fig:powder_pcc_contribution_best_w_binning}
    }
    \hfill
    \subfloat[]{
        \includegraphics[width=0.45\textwidth]
        {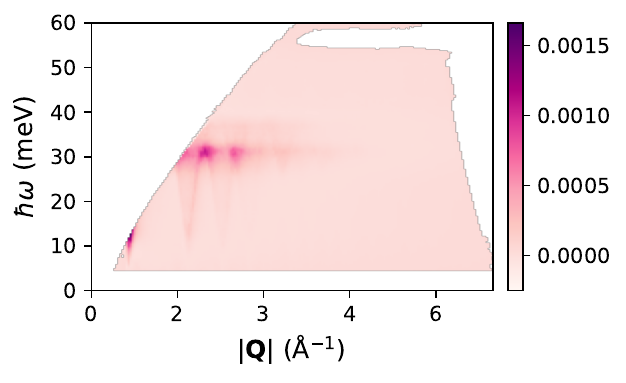}
        \label{fig:powder_pcc_contribution_ref_w_binning}
    }
    \caption{
    Pixel-wise decomposition of the powder spectra Pearson correlation coefficient, $\mathrm{PCC}_{\mathrm{pd}}$, for
    \textbf{(a)} the selected Hamiltonian $\bm{\theta}_{\mathrm{best}}$ obtained with finite $\mathbf{Q}$-space binning and
    \textbf{(b)} the literature reference Hamiltonian $\bm{\theta}_{\mathrm{ref}}$.
    }
    \label{fig:powder_pcc_contribution_w_binning}
\end{figure}

\begin{figure}[htbp]
    \centering
    \subfloat[]{
        \includegraphics[width=0.45\textwidth]
        {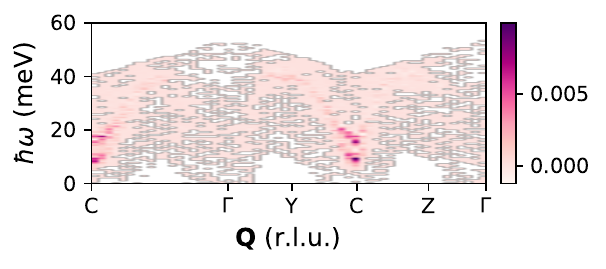}
        \label{fig:sc_pcc_contribution_best_w_binning}
    }
    \hfill
    \subfloat[]{
        \includegraphics[width=0.45\textwidth]
        {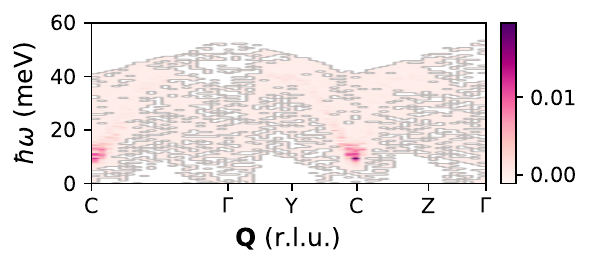}
        \label{fig:sc_pcc_contribution_ref_w_binning}
    }
    \caption{
    Pixel-wise decomposition of the single-crystal spectra Pearson correlation coefficient, $\mathrm{PCC}_{\mathrm{sc}}$, for
    \textbf{(a)} the selected Hamiltonian $\bm{\theta}_{\mathrm{best}}$ obtained with finite $\mathbf{Q}$-space binning and
    \textbf{(b)} the literature reference Hamiltonian $\bm{\theta}_{\mathrm{ref}}$.
    }
    \label{fig:sc_pcc_contribution_w_binning}
\end{figure}

Figure~\ref{fig:expt_conerplot_spectrograms_w_binning} presents the resulting finite-bin-averaged spectral comparison, while Figs.~\ref{fig:powder_pcc_contribution_w_binning} and~\ref{fig:sc_pcc_contribution_w_binning} show the corresponding pixel-wise PCC contribution maps. The similar overall conclusions obtained with and without finite-bin averaging indicate that the reported multimodal spectral comparison is robust to this difference in representation. Table~\ref{tab:hamiltonian_parameters_binned} lists the inferred Hamiltonian parameter values.

% Best param: [-8.21811613e-03  2.09885821e-01 -2.54840446e+00 -2.09825182e+00  2.08364114e-01  1.74457043e-01  1.37762108e+01  1.48604393e+01 -4.92412567e-01]
\begin{table}[htbp]
\caption{
Inferred Hamiltonian parameters obtained with finite $\mathbf{Q}$-space averaging (meV).
}
\label{tab:hamiltonian_parameters_binned}
\centering
\begin{tabular}{lccccccccc}
\hline
 & $A_x$ & $A_z$ &
 $J_{\mathrm{1a}}$ & $J_{\mathrm{1b}}$ &
 $J_{\mathrm{2a}}$ & $J_{\mathrm{2b}}$ &
 $J_{\mathrm{3a}}$ & $J_{\mathrm{3b}}$ &
 $J_{\mathrm{4}}$ \\
\hline
$\bm{\theta}_{\mathrm{best,binned}}$
&
$-0.0082$
&
$0.21$
&
$-2.55$
&
$-2.10$
&
$0.21$
&
$0.17$
&
$13.78$
&
$14.86$
&
$-0.49$
\\
\hline
\end{tabular}
\end{table}

\clearpage
\subsection{Extraction of powder magnon spectrum from the experimental INS data}
\label{sec_SN:extraction_expt_powder_data}

\begin{figure}[htbp]
    \centering
    \includegraphics[width=0.75\linewidth]{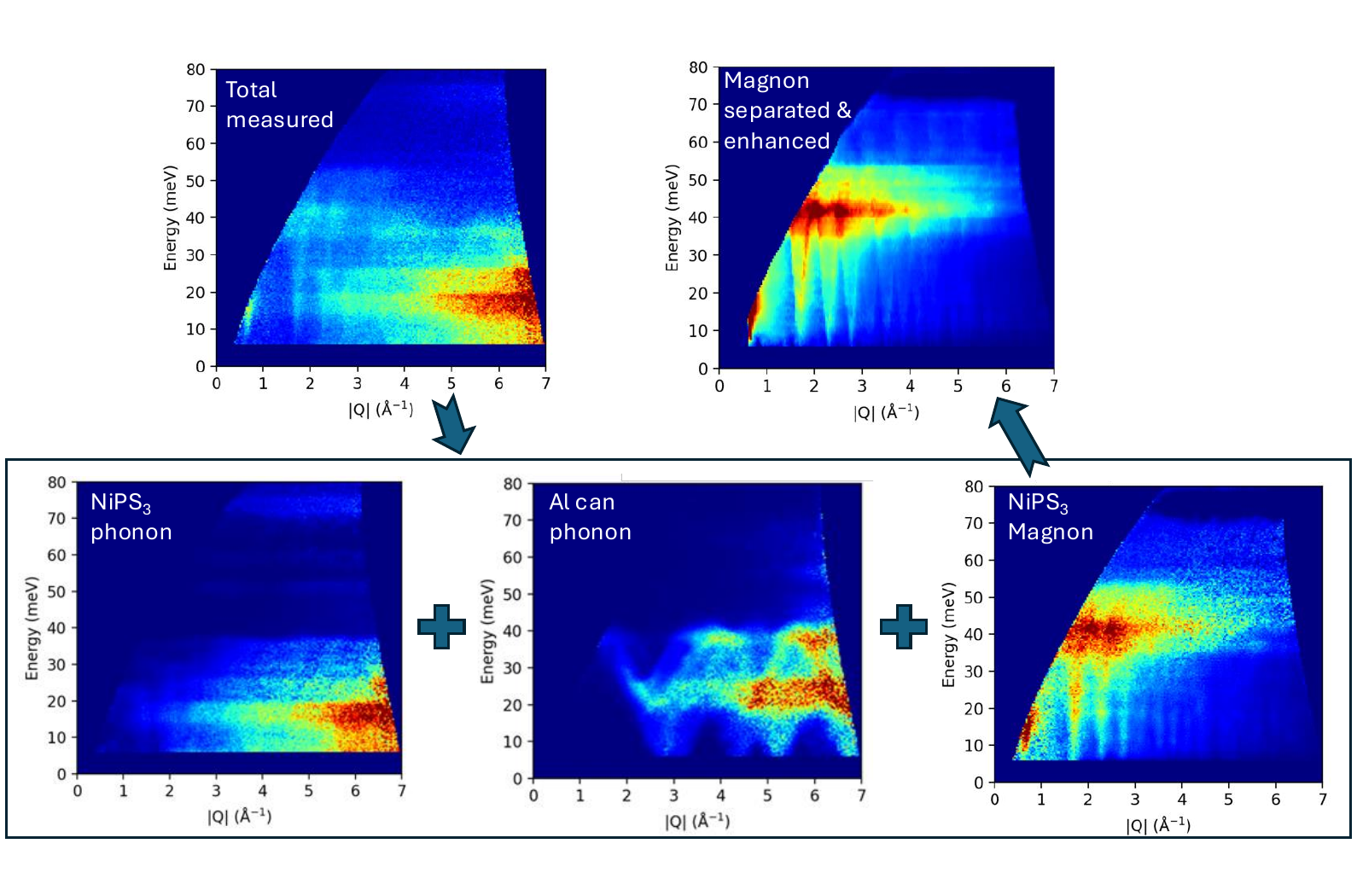}
    \caption{Source separation of the measured INS spectrum into three components: sample magnons, sample phonons, and phonons originating from the aluminum sample holder. Each panel uses an independent color scale to better visualize the spectral features of the corresponding component. On a common intensity scale, the measured total spectrum is the sum of these three contributions. The isolated magnon spectrum is subsequently processed with a feature-enhancement model to suppress noise and sharpen physically consistent spectral features. The elastic peak and its low-energy tail near $E = 0$ have been masked.}
    \label{fig:SI_powder_separation}
\end{figure}

INS measures not only magnetic excitations (magnons) but also lattice excitations (phonons) from the sample. In addition, the nonmagnetic aluminum sample holder contributes phonon scattering to the measured spectrum. Magnon scattering is typically strongest at low momentum transfer ($|\mathbf{Q}|$) and decreases with increasing $|\mathbf{Q}|$ due to the magnetic form factor. In contrast, phonon scattering generally becomes more prominent at higher $|\mathbf{Q}|$. In orientationally averaged powder spectra, the magnon and phonon contributions partially overlap. In this case, although most magnon intensity is concentrated at $|\mathbf{Q}|<3\,\mathrm{\AA}^{-1}$ and most phonon intensity appears at $|\mathbf{Q}|>3\,\mathrm{\AA}^{-1}$, the two contributions still overlap, and their separation remains essential because the AI surrogate model predicts only magnon spectra.
To isolate the magnetic signal, we trained a source-separation vision transformer using synthetic datasets that include a variety of NiPS\textsubscript{3} magnon spectra, NiPS\textsubscript{3} phonon spectra, and aluminum phonon spectra. The extracted magnon spectrum was subsequently denoised using a feature-enhancement vision transformer to improve signal quality prior to inference. The spectra that go through the data processing workflow are illustrated in Fig.~\ref{fig:SI_powder_separation}.

As shown in Fig.~\ref{fig:SI_powder_separation}, the final NiPS3 magnon spectrum preserves the key features of the measured magnetic signal. The source separation effectively isolates the magnon contribution from the phonon background, while the feature-enhancement model suppresses noise and sharpens spectral features within the constraints imposed by the underlying physical model. A detailed description of the source separation and spectral processing workflow using vision transformers will be presented in a separate publication, as its methodology is beyond the scope of the present work.

Because the source-separation and feature-enhancement procedures introduce additional model-dependent processing beyond the raw experimental measurement, the resulting powder spectrum should be interpreted as an effective magnetic-spectrum estimate rather than a model-independent experimental observable. The experimental powder analysis presented here is intended primarily to demonstrate the multimodal inference workflow; quantitative differences in the inferred Hamiltonian may depend on the specific preprocessing procedure. A systematic comparison of alternative powder-data processing strategies is left for future work.

\end{document}